\documentclass[A4paper,useAMS,usenatbib]{mn2e}
\usepackage{graphicx,amssymb,color}
\usepackage{lastpage}
\usepackage[fleqn]{amsmath}
\usepackage[normalem]{ulem}
\input psfig

\title[Full-life sims of unequal-mass planets]
{Full-lifetime simulations of multiple unequal-mass planets across all phases of stellar evolution}

\author[Veras et al.]
{Dimitri Veras$^{1}$\thanks{E-mail: d.veras@warwick.ac.uk},
Alexander J. Mustill$^{2}$,
Boris T. G\"{a}nsicke$^{1}$,
Seth Redfield$^{3}$,
\newauthor
Nikolaos Georgakarakos$^{4}$,
Alex B. Bowler$^{1}$,
Maximillian J.S. Lloyd$^{1}$
\\
$^{1}$Department of Physics, University of Warwick, Coventry CV4 7AL, UK
\\
$^{2}$Lund Observatory, Department of Astronomy and Theoretical Physics, Lund University, Box 43, SE-221 00 Lund, Sweden
\\
$^{3}$Astronomy Department and Van Vleck Observatory, Wesleyan University, Middletown, CT 06459-0123, USA
\\
$^{4}$New York University Abu Dhabi, Saadiyat Island, P.O. Box 129188, Abu Dhabi, UAE
}

\date{Accepted 2016 February 25. Received 2016 February 25; in original form 2016 January 27}
\pubyear{2016}

\begin{document}
\label{firstpage}
\pagerange{\pageref{firstpage}--\pageref{lastpage}}
\maketitle

\begin{abstract}
We know that planetary systems are just as common around white dwarfs as around main sequence stars. However, self-consistently linking a planetary system across these two phases of stellar evolution through the violent giant branch poses computational challenges, and previous studies restricted architectures to equal-mass planets. Here, we remove this constraint and perform over 450 numerical integrations over a Hubble time (14 Gyr) of packed planetary systems with unequal-mass planets. We characterize the resulting trends as a function of planet order and mass. We find that intrusive radial incursions in the vicinity of the white dwarf become less likely as the dispersion amongst planet masses increases.  The orbital meandering which may sustain a sufficiently dynamic environment around a white dwarf to explain observations is more dependent on the presence of terrestrial-mass planets than any variation in planetary mass.  Triggering unpacking or instability during the white dwarf phase is comparably easy for systems of unequal-mass planets and systems of equal-mass planets; instabilities during the giant branch phase remain rare and require fine-tuning of initial conditions.  We list the key dynamical features of each simulation individually as a potential guide for upcoming discoveries.
\end{abstract}


\section{Introduction}

Nearly 100 planets are known to orbit giant stars \footnote{www.lsw.uni-heidelberg.de/users/sreffert/giantplanets.html}, and signatures of planetary systems have been detected at over 1000 white dwarfs. This latter number is obtained through observed planetary debris in white dwarf atmospheres \citep{zucetal2003,dufetal2007,zucetal2010,kleetal2013,koeetal2014,genetal2015,kepetal2015,kepetal2016}. About 40 of these white dwarfs contain compact ($\approx 0.6-1.2R_{\odot}$) planetary debris discs (see \citealt*{farihi2016} for a review), and one hosts at least six transiting planetesimals (WD 1145+017: \citealt*{vanetal2015,croetal2016,gaeetal2016,rapetal2016,xuetal2016}).  Also, planets around two other white dwarfs have been observed (WD 0806-661 b: \citealt*{luhetal2011}, and PSR B1620-26AB~b: \citealt*{sigetal2003}).  Although the architectures of most white dwarf planetary systems remain unknown, these statistics demonstrate that the study of post-main-sequence planetary systems has entered a new era, one where we can begin to investigate population-wide trends as well as key individual systems.  $N$-body simulations of multi-planet systems represent a vital probe into their history and future, revealing insights about their formation and fate.

However, accurately performing multi-body simulations across different phases of stellar evolution remains challenging. For bodies much smaller than planets, including gravity alone is likely to be insufficient \citep[see Fig. 2 of][]{veras2016}.  Asteroids within about 7 au of a main sequence star could be spun up to fission during the giant branch phase of stellar evolution \citep{veretal2014a} due to intense radiation.  Asteroids further away could have their orbits changed due to another radiation-based effect: the Yarkovsky drift \citep{veretal2015a}.  Further, the stellar wind could induce drag on asteroids and pebbles \citep{donetal2010,veretal2015a}, and sublimation of volatile substances on these objects could change their orbits \citep{veretal2015b} and/or launch ejecta, as speculated in WD 1145+017 \citep{vanetal2015,croetal2016,gaeetal2016,rapetal2016,xuetal2016}.

Even restricting simulations to planets presents challenges: (i) Tidal effects between planets and their parent stars can destroy or alter the planets, but just how remains an open question (see Sec. 5 of \citealt*{veras2016} for a review). Giant branch (GB) stars harbour radii that extend to several au, and planets too close to their parent stars may hence be engulfed on both the red giant branch (RGB) phase \citep[e.g.][]{viletal2014} and the asymptotic giant branch (AGB) phase \citep[][]{musvil2012}. Nevertheless, only about half of the currently known exoplanets will likely be engulfed \citep{norspi2013}, and observational biases against finding planets at large separations imply that the actual fraction is much less. (ii) Computational limitations hinder explorations with long main-sequence lifetimes or planets on close-in orbits. Only recently \citep{vergae2015} have $14$ Gyr (the current age of the Universe) simulations with main sequence progenitor masses under $3M_{\odot}$ (most white dwarf progenitors had masses between about $1.5M_{\odot}$ and $2.5M_{\odot}$; \citealt*{koeetal2014}) been carried out for ensembles of multi-planet systems\footnote{A few individual planetary systems, or putative planetary systems, have been modelled with simulations spanning multiple phases of evolution: HU Aqr \citep{portegieszwart2013} and NN Ser \citep{musetal2013}.}, as previous attempts \citep{dunlis1998,debsig2002,veretal2013a,musetal2014} did not achieve this coverage.

Nevertheless, up until now full-lifetime simulations of multi-planet systems have been restricted to equal-mass planets. Although this assumption significantly helps constrain the available parameter space to explore, real systems exhibit a variance of planetary masses of a few percent to many orders of magnitude. 
Further, previous studies have predominately modelled Jupiter-mass planets, which are rarer than terrestrial planets \citep{casetal2012,winfab2015}. Further, no published study has simulated multiple planets with test particles.

Here, we break these barriers, and perform a suite of 14 Gyr simulations of unequal-mass planets, occasionally including test particles, in order to explore the consequences and resulting trends.  
In Section 2, we describe our setup. Section 3 details the classification scheme for our results, and the results themselves.  We discuss the implications in Section 4, and conclude in Section 5.

Appendix A is our simulation database. Each row of each table corresponds to one simulation, and within each row we present the salient dynamical features.

\section{Simulation setup}

Simulations of planetary systems through multiple stages of stellar evolution require both the star and planets to be treated self-consistently as a function of time.

\subsection{Numerical codes}

Here we have used an updated version of the code from \cite{veretal2013a}, \cite{musetal2014}, \cite{veretal2014b} 
and \cite{vergae2015}, which combines planetary and stellar evolution. The stellar evolution is computed from {\sc SSE} \citep{huretal2000},
which is more than sufficiently accurate for our purposes.  If we instead desired to trace more detailed
characteristics of a particular star, like its chemical profile, then perhaps a code like the increasingly utilized
MESA \citep{paxetal2011,paxetal2015} would be more suitable. However, here we need only the mass and radius evolution 
of the star, and did not model any particular known system; we ignored radiative effects, which are negligible 
for the types of planets we simulated.

The output from {\sc SSE} was ported directly into a heavily modified version of the {\sc Mercury}
planetary evolution code, originally from \cite{chambers1999}. Our version of {\sc Mercury} used 
the Bulirsch-Stoer integrator throughout the simulation,
ensuring accurate treatment of potential close encounters. We adopted a tolerance value of $10^{-13}$. 
Stellar mass and radius changes were interpolated
within each Bulirsch-Stoer timestep, helping to ensure accuracy. Stars which engulfed planets
throughout the course of the simulations had masses which were increased accordingly.  Our
output frequency was 1 Myr; a shorter frequency would have prohibitively slowed down our simulations.
As is the {\sc Mercury} default, any collisions between planets were treated as purely inelastic.
Further, our modified code allowed for the tracking of the minimum orbital pericentre of all
surviving planets, and adopted a standard Hill ellipsoid for the solar neighbourhood \citep{vereva2013,veretal2014c}
to accurately track ejections.

\begin{figure*}
\centerline{
\psfig{figure=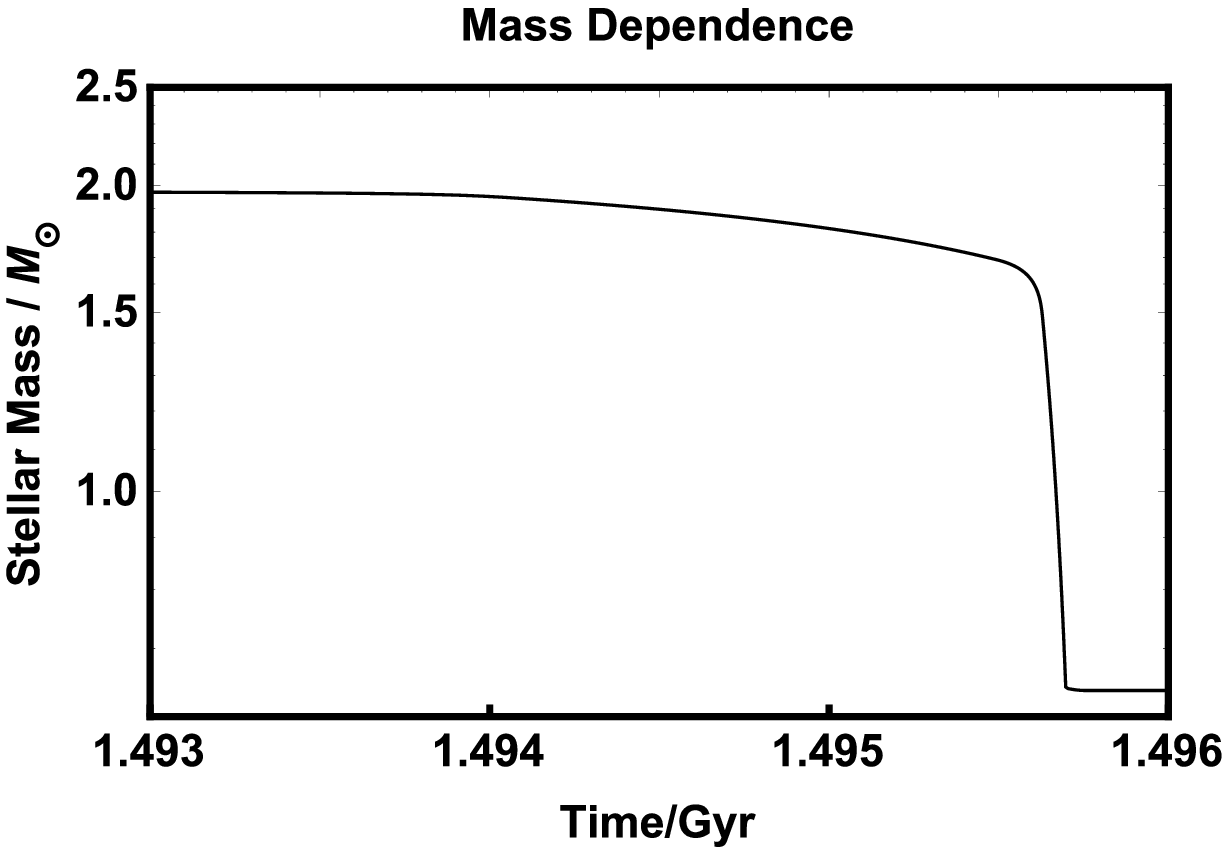,width=8cm,height=6cm}
\ \ \ \ \ \ \ \
\psfig{figure=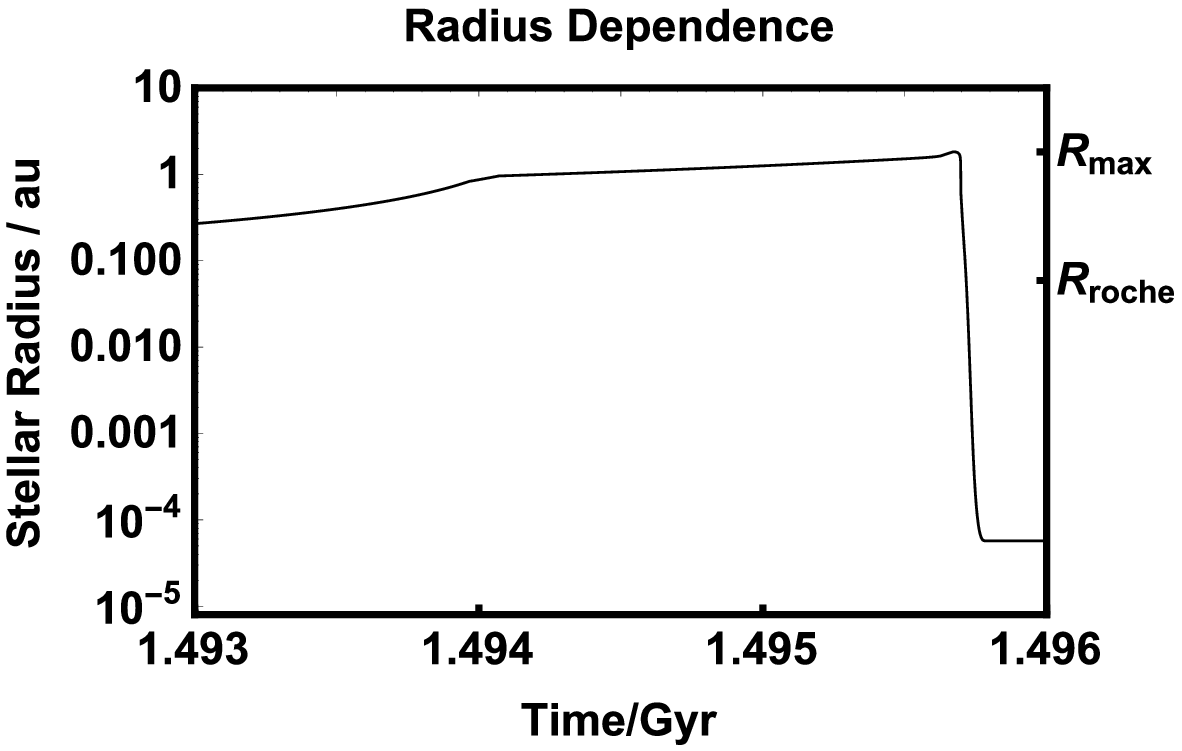,width=10cm,height=6cm}
}
\caption{
The mass ({\it left panel}) and radius ({\it right panel}) 
evolution of the star used in this study, during the tip of 
the AGB phase and when the white dwarf is born (at the 
start of the EWD= ``early white dwarf'' phase).  Marked on the right panel is
the maximum AGB radius ($R_{\rm max}$) and the white dwarf
Roche radius (distance) adopted in this study ($R_{\rm roche}$).
}
\label{steevo}
\end{figure*}

\subsection{Stellar properties}

The enormous parameter space of our computationally-demanding simulations
forced us to adopt a single type of star for our simulations. 
Our star contained a physically-motivated stellar mass of 
$2.0M_{\odot}$ on the main sequence.  The present-day population of white dwarfs,
with average masses ranging from about $0.60 M_{\odot}$ to $0.65 M_{\odot}$ \citep{lieetal2005,faletal2010,treetal2013} 
corresponds to main sequence A- and F-star progenitors (see Fig. 3 of \citealt*{veras2016}), from which $2.0M_{\odot}$ is
an appropriate value from the initial-to-final mass relation \citep{catetal2008,kaletal2008,casetal2009,koeetal2014}. 
This value coincidentally also marks (i) the point beyond which the planet occurrence rate falls off \citep{refetal2015}
and (ii) a transition in evolutionary sequence due to stellar mass; below $2.0M_{\odot}$ a star would continue ascending 
the RGB until it undergoes a core helium flash, which changes the amount of mass lost and radius along the RGB. Lower-mass
stars have larger radii and greater mass loss. Regardless, the
greatest mass loss (by several orders of magnitude) and radius changes occur along the AGB even for values within
a few $0.1M_{\odot}$ of 2.0$M_{\odot}$.

The evolution of the star is illustrated
in Fig. \ref{steevo}, characterized in Table \ref{EvolutionTable}
and described in this paragraph.
Our $2.0 M_{\odot}$ star was assumed to have Solar metallicity, and 
remained on the main sequence for 1.1735 Gyr.
Along the RGB, the star lost mass according to the 
traditional\footnote{\cite{schcun2005} provided an improved, physically-motivated
version of this prescription, but one that requires knowledge  
of further details (surface gravity, temperature) about the star.}
Reimers mass loss prescription, with a numerical coefficient of 0.5.  The star
stayed on the RGB for 23 Myr and lost $0.002 M_{\odot}$ during that time, while
expanding its radius out to 0.13 au. Afterwards, the star contracted
to a radius of 0.04 au.  The AGB phase began at 1.4894 Gyr,
and lasted for only 6.4 Myr. However, during this time, the star
lost $1.338 M_{\odot}$ and expanded its radius out to $1.82$ au;
see Fig. \ref{steevo}. 
Finally, the star ended its evolution as a white dwarf formed with a 
mass of $0.6365M_{\odot}$ and a radius of $5 \times 10^{-5}$ au.

\begin{table}
\caption{Time at the beginning of each phase 
(MS $=$ main sequence, 
 GB $=$ giant branch,
EWD $=$ early white dwarf,
LWD $=$ late white dwarf) and the total
mass lost during those phases. The LWD 
phase lasts until the end of the simulations (14 Gyr).}
\begin{tabular}{| c | c  c  c  c}
  & MS & GB & EWD & LWD \\
\hline
Start Time (Gyr)              & 0.0 & 1.1735 & 1.4958 & 1.5958 \\
Mass Lost ($M_{\odot}$)  & 0.0 & 1.363  & 0.0  & 0.0 \\
\end{tabular}
\label{EvolutionTable}
\end{table}

Varying the stellar mass within the code is nearly equivalent to assuming that the star
loses mass isotropically.  This assumption is excellent for orbiting bodies 
within a few hundred au \citep{veretal2013b}.  Because the planets are assumed to be point
masses, they do not accrete any of the stellar mass and the isotropic assumption
is maintained; see Sec. 4 of \cite{veras2016} for more details. This type of stellar mass decrease, 
however, does not consider the lag time between the ejecta passing beyond two different
orbits.  However, this effect should be negligible; see section 2 of
\cite{payetal2016} for quantification.

A planet that ventures into the vicinity of the white dwarf might
be disrupted or destroyed. This ``vicinity'' may extend to a few hundred
times the white dwarf radius. The critical radius at which disruption occurs
(known as the Roche radius),
however, is dependent on the planet's shape, composition, spin state, 
orbital state, and whether one considers disruption to mean cracking,
deforming or dissociating.  This ambiguity
is compounded by the fact that no study has yet modelled the disruption 
of a planet around a white dwarf\footnote{Main-sequence disruption investigations 
\citep{guietal2011,liuetal2013} suggest that the assumed structure of the 
planet plays a vital role, as well as how much mass is sheared off
during each close passage to the star.}. Although the disruption of rubble pile 
asteroids around white dwarfs has been numerically modelled \citep{debetal2012,veretal2014d},
the situation with planets is fundamentally different.  These uncertainties prompted
us to rescale the white dwarf radius within the simulations to a value corresponding
to its fiducial Roche, or disruption, radius: $1.27R_{\odot} \approx 0.0059$ au, 
where $R_{\odot}$ is the Sun's radius.
This value roughly represents the outer extent of the compact debris discs
which surround white dwarfs (see \citealt*{farihi2016} for a review). These
discs are assumed to be composed of disrupted fragments and particles.

\subsection{Planet properties}

Our goal is to simulate planetary systems that become unstable.
Instability in planetary systems is likely to be common, as demonstrated by the Grand Tack model
\citep{pieray2011,waletal2011,obretal2014,izietal2015} and the Nice model 
\citep{gometal2005,moretal2005,tsietal2005,levetal2011} for our solar system, and
by the potential future instability of packed exoplanetary systems, 
which are prevalent (recent examples include \citealt*{baretal2015}
and \citealt*{cametal2015}; see also 
\citealt*{puwu2015})\footnote{Rarely has the future non-secular evolution of planetary systems
throughout the entire main sequence been achieved with $N$-body numerical integrations. Consequently, the prospects for future instability
of the currently-observed exoplanetary systems is generally unknown.}.  Further, metal-polluted white dwarfs,
which comprise between one-quarter and one-half of all Milky Way white dwarfs \citep{zucetal2003,zucetal2010,koeetal2014}, 
are thought to arise from planetary system instability after the star has become a white dwarf.

We consider simulation suites of primarily 4-planet systems in order to facilitate comparison
with the equal-mass cases of \cite{vergae2015}, although we also ran smaller
samples of six- and eight-planet systems.  We also adopted simulations that each contained four planets
and 12 test particles. Each test particle represents a planet or asteroid which
is both (i) small enough relative to the nonzero-mass planets to not affect them, 
and (ii) large enough to not be affected by radiation, which is not modelled.  
One example is four giant planets with test particles represented by Earths.
Large asteroids with radii above about 100 km may also be represented as test particles,
because the effect of radiation for objects of these sizes may be negligible 
(see eqs. 108 and 110 of \citealt*{veretal2015a}
and eq. 1 of \citealt*{veretal2014a}).

For our nonzero-mass substellar bodies, we adopted eight types of planets: Jupiter, Saturn, Uranus, Neptune (which we refer to as ``giant planets''),
and all of their analogues scaled down in mass by a ratio of $M_{\rm Jupiter}/M_{\oplus} \approx 317.8$ (which we refer to as ``terrestrial planets''). The mass scaling effectively transforms Jupiter into Earth, and the other giant planets into three sub-Earth mass companions.  The scaled-down planets allow us to provide direct dynamical comparisons while keeping the mass ratios amongst the planets the same. These terrestrial planets also arguably yielded the most interesting results. We adopted giant planet densities which reflect those of Jupiter, Saturn, Uranus and Neptune.  The density of all the scaled-down planets was set to the density of the Earth.

We henceforth denote the giant planets as {\tt J}, {\tt S},  {\tt U},  {\tt N}, and the terrestrial planets as {\tt \={J}}, {\tt \={S}}, {\tt \={U}}, and {\tt \={N}}. 
All of the planetary system combinations that 
we adopted per simulation are presented in the Appendix. For perspective on the relative mass values, see 
Table \ref{MassTable}.  Test particles are by defintion massless, and can reasonably represent objects (whether they be planets, asteroids or pebbles) which are at least two or three orders of magnitude less massive than the non-zero mass bodies in the simulation.  Also, $M_{\rm Jupiter}/M_{\odot} = 9.54 \times 10^{-4}$ and $M_{\oplus}/M_{\odot} = 3.00 \times 10^{-6}$.

\begin{table}
\caption{Mass ratios of different planets 
({\tt J} $=$ Jupiter, 
 {\tt S} $=$ Saturn,
 {\tt N} $=$ Neptune,
 {\tt U} $=$ Uranus). These ratios are equivalent
to those of the planets which are scaled-down in mass 
(\tt \={J}, \tt \={S}, \tt \={N}, \tt \={U}).}
\begin{tabular}{| c | c  c  c  c}
  & {\tt J} & {\tt S} & {\tt N} & {\tt U} \\
\hline
{\tt J} & 1     & 0.30 & 0.054 & 0.046 \\
{\tt S} & 3.34  & 1    & 0.18  & 0.15  \\
{\tt N} & 18.53 & 5.55 & 1     & 0.85  \\
{\tt U} & 21.87 & 6.55 & 1.18  & 1     \\
\end{tabular}
\label{MassTable}
\end{table}

Our choices for initial orbital eccentricities, inclinations, orbital angles, and innermost semimajor axis
follow those of previous studies \citep{musetal2014,vergae2015} and their justifications are only briefly 
repeated here.  All planets are assumed to be on initially circular orbits and have small inclinations 
randomly selected from a uniform distribution from $-1^{\circ}$ to $1^{\circ}$. Adopting strictly non-coplanar planets
prevent an unnaturally high rate of planet-planet collisions, which occurred in \cite{veretal2013a}. Imposing
non-zero initial eccentricities would change (speed up) instability timescales; we did not do so in order
to facilitate comparisons with previous studies. The innermost planet semimajor axis was always set at 5 au to prevent AGB
star-planet tides from playing a role in the evolution \citep[see Fig. 7 of][]{musvil2012} before
any potential instability occurs.  Further, 5 au is a particularly appropriate value considering that Jupiter lies 
at 5.2 au from the Sun and is the closest of the four giant planets in our Solar system.

The much trickier initial parameter to determine was the initial spacing of the planets. 
For equal-mass planets, the link with initial spacing and instability timescales has a
now-substantial history (see \citealt*{davetal2014} for a review), particularly with the application
of the mutual Hill radius as the separation unit.  However, no widely-used formalism
exists with unequal-mass planets. Consequently, for lack of better proven alternatives, we applied
the mutual Hill radius to our architectures here. Multiple definitions of this parameter exist:
we adopted eq. (4) of \cite{smilis2009} in order to maintain consistency and provide meaningful
comparisons with \cite{vergae2015}:

\begin{eqnarray}
a_{i+1} &=& a_i 
\left[1 + \frac{\beta}{2}\left( \frac{m_i + m_{i+1}}{3 \left(m_{\star} + \sum_{k=1}^{i-1}m_{k} \right)}  \right)^{1/3}\right]
\nonumber
\\
&\times&
\left[1 - \frac{\beta}{2}\left( \frac{m_i + m_{i+1}}{3 \left(m_{\star} + \sum_{k=1}^{i-1}m_{k} \right)}  \right)^{1/3}\right]^{-1}
.
\label{theequation}
\end{eqnarray}

\noindent{}In this equation, $a$ and $m$ refer to mass and initial semimajor axis, and the subscripts ascend
in order of increasing distance from the star. The important quantity $\beta$ is the number of mutual Hill radii.
In order to determine meaningful values of $\beta$ for the different architectures we considered, 
we performed exploratory preliminary suites of simulations.  
We found that a wide range ($\beta = 6-14$) was necessary to implement depending on the architecture
considered.  The specific values used for each architecture are listed in Tables \ref{JUUU}-\ref{UNUN}.

Having established the planet locations, we then considered where potential test particles would reside. We distributed our 12 test particles uniformly in a ring at 2.5 au from the star.  This choice is in the spirit, if not the
details, of the asteroid belt. In our Solar system, the largest objects in this belt (with sizes greater
than 100 km) are unlikely to be influenced by Solar giant branch radiation, and will neither be engulfed
by the Solar giant. Recall that these particles could instead represent Mars, which also will survive
the Sun's post-main-sequence evolution, despite being located at about 1.5 au \citep{schcon2008}.

\subsection{Additional physics}

Besides radiation, other physics that could play a role in planetary system evolution
include star-planet tides and general relativity. A planet which is perturbed on an
orbit with a pericentre that lies just outside of the Roche radius may be tidally circularized.
The particulars of this process are highly dependent on the composition of the approaching
planet and the evolutionary stage of the star; all our bodies are point-masses with no assumed composition. 
The variation in tidal
circularization behaviour and timescale due to composition is so great \citep{henhur2014} that
any meaningful exploration would require a dedicated study, which we do not perform here. 
Our simulations here illustrate preconditions for this tidal interaction to occur.

General relativity changes the rate of the argument of periastron for close-in bodies, 
and hence can by itself trigger instability in multi-planet systems \citep[e.g.][]{verfor2010}.
Consequently, we have included the effects of general relativity in our simulations through
our updated code.

\subsection{Running time}

We attempted to run all our simulations for 14 Gyr, which represents a Hubble time and is the
current age of the Universe.  We succeeded in over 90\% of cases, the exceptions (which are all noted in the Appendix tables) being systems where a planet or test particle was perturbed close 
enough to the star to sufficiently slow
down the simulations.  We only report simulations which ran for at least 1.9 Gyr (recall that our star
becomes a white dwarf after about 1.5 Gyr), in order to give a flavour
of what the evolution is like on all of the main sequence, giant branch, and early white dwarf phases.

\section{Results}  \label{secRes}

We present results for over 450 simulations, and have visually inspected the output and evolution of each one.
They are partitioned into groups of up to four simulations such that each group member has the same (i) initial ordering
and type of masses (such as {\tt \=N\=U\=J\=S} in order of increasing distance from the star), and (ii) value of $\beta$. 
Within these groups the initial orbital angles and inclinations are different. 

We report the results of every simulation in Tables \ref{JUUU}-\ref{UNUN}, one per row,
with particular attention to the stellar phases at which various events occurred rather
than the specific times.  This format allows one to determine qualitative trends easily amongst 
the many-dimensional parameter space, and acts as a handy reference for setting up future simulations
if one has a desired outcome or set of initial conditions (perhaps based on a known exosystem) in mind.

In this section, we describe the data which is presented in the tables (Subsection \ref{tabcol}), 
illustrate some representative and interesting examples (Subsection \ref{specas}),
list various system outcomes and behaviours which our simulations show to be possible (Subsection \ref{sysbeh}), 
and analyze the general trends
from the tables (Subsection \ref{gentre}).


\subsection{Description of table columns}  \label{tabcol}

In all tables, the first column (``Sim \#'') provides a designation for each simulation for easy reference.
The second column (``Setup'') provides the initial order and type of planets in each simulation.
We reiterate that {\tt J}, {\tt S},  {\tt U},  {\tt N}, {\tt \={J}}, {\tt \={S}}, {\tt \={U}}, and {\tt \={N}},
respectively refer to planets which have the same masses of Jupiter, Saturn, Uranus, Neptune, and versions
of those planets with masses scaled down by a factor of about 318 (thereby transforming Jupiter into Earth).  
All simulations in Tables \ref{JUUU}-\ref{JUNS} contain
four planets each.  The last two tables (\ref{JSJS}-\ref{UNUN}) contain systems with four, six and eight planets.
The third column (``$\beta$'') refers to the number of mutual Hill radii between the planets, as defined from
eq. (4) of \cite{smilis2009}.

Starting from the fourth column (``Unpack'') we characterize the timing of events in the evolution of the 
planetary system.  We adopt designations for different phases of stellar evolution:
MS (main sequence), GB (giant branch), EWD (``early white dwarf'' that corresponds to stars that
have become white dwarfs within the last 100 Myr), and LWD (``late white dwarf'' stars which became white dwarfs over 100 Myr ago).
We split up the white dwarf phase because the intense mass loss at the tip of the AGB phase often triggers
slightly-delayed instability, which commonly manifests itself in white dwarfs whose cooling age (the time since becoming a white dwarf)
is less than about 100 Myr.  In effect, such systems are dynamically reseting themselves and hence feature
instability at ``early'' times, just as we would expect from a planetary system recently born out of a Solar nebula.
Precisely then, the MS phase corresponds to times between 0 and 1173.576 Myr, while the GB phase corresponds to times between
1173.576 Myr and 1495.783 Myr.  At 1495.783 Myr, the EWD phase begins and lasts until 1595.783 Myr. The star then spends the remainder
of its life on the LWD phase.

The fourth column itself (``Unpack'') displays the phase during which the system became unpacked. If the
system never became unpacked, then the space is left blank. We define ``unpacked'' as the moment that either
(i) two non-zero mass planets cross orbits, or (ii) an instability occurs.  We define instability as an occurrence when
two bodies collide with one another, or one body escapes the system.  The collision could come in the form of
a star-planet collision (when the planet is said to be engulfed in the star) or a planet-planet collision. Note
also that the moment of escape may occur several Myr after the actual interaction which triggered the movement,
because the Hill ellipsoid of the system typically lies at about $10^5$ au from the star.

The fifth column (``\# Surv'') indicates the number of non-zero mass planets which remained in the system
by the end of the simulation. Those planets which do not survive are characterized in the next three columns 
(``Engulf'', ``Eject'' and ``Collision''), which indicate respectively when a planet intersects with the
stellar radius, is ejected from the system, or hits another planet.   Recall that the white dwarf stellar radius
is enhanced from its true value. The columns all indicate the phase in which an instability occurred,
along with the planet(s) involved in the instability in the subscripts. The subscript numbers correspond to the planet order from the ``Setup'' column. 
Each instability is indicated by a 
single listed entry. The subscripts in each ``Collision'' entry indicate the two planets involved in the collision.

The column (``$< R_{\rm max}$'') lists any non-zero mass planet that survived for the entire simulation
and was perturbed into an orbit along the EWD or LWD phase whose pericentre was within the star's maximum 
AGB radius (1.82 au). The subscript indicates the smallest planet-white dwarf distance achieved.

The column, labelled ``TPs Eng'', does not exist in the final two tables (\ref{JSJS}-\ref{UNUN}).
The column indicates when test particles were included in the simulations (a blank space means no test particles), 
and provides some information about them. The first and second numbers given are the amounts of test particles (out of 12) that were
engulfed by the star during the EWD and LWD phases, respectively.  

The final column lists relevant notes which are in the
table captions.

\subsection{Specific cases}   \label{specas}

Now we present some specific examples of evolutionary sequences.

\subsubsection{Standard giant planet evolution}

\begin{figure}
\psfig{figure=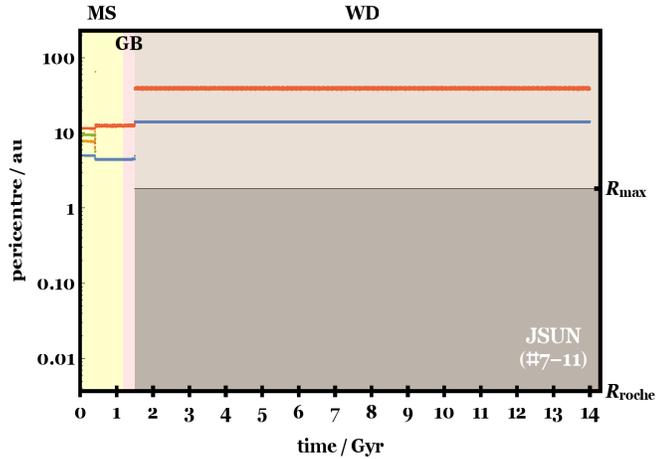,width=9cm,height=6.5cm}
\caption{
A characteristic outcome for the full-lifetime evolution of four giant planets:
unpacking and instability on the main sequence, followed by stability. Note that
the surviving two planets expand their orbits due to giant branch mass loss at 1.49 Gyr.
The values of $R_{\rm max}$ and $R_{\rm roche}$ on the right axis indicate the maximum
stellar AGB radius (1.82 au) and an approximate value of the disruption radius of
white dwarf (0.0059 au).  Shown is {\tt JUUU} simulation \#1-19 (Table \ref{JUUU}). 
}
\label{sym1-19}
\end{figure}

Consider first simulation \#1-19 (in Table \ref{JUUU}), whose evolution is shown in Fig. \ref{sym1-19}.
The system initially consists of an inner Jupiter-mass planet (blue, at 5 au) followed
by three Uranus-mass planets ({\tt JUUU}), separated by $\beta = 8$. The system unpacks on the main sequence, 
and both the second and third planets (Uranuses) are ejected sometime during this phase. 
The two remaining planets have their orbits expanded due to mass loss at the end of the GB
phase, remaining stable through this process and for the remainder of the simulation.  Neither achieved an 
orbit that took it to within 1.82 au ($= R_{\rm max}$) of the white dwarf, and their 
pericentres remain nearly constant.

\subsubsection{Squeezed solar system analogue -- giant planets}

\begin{figure}
\psfig{figure=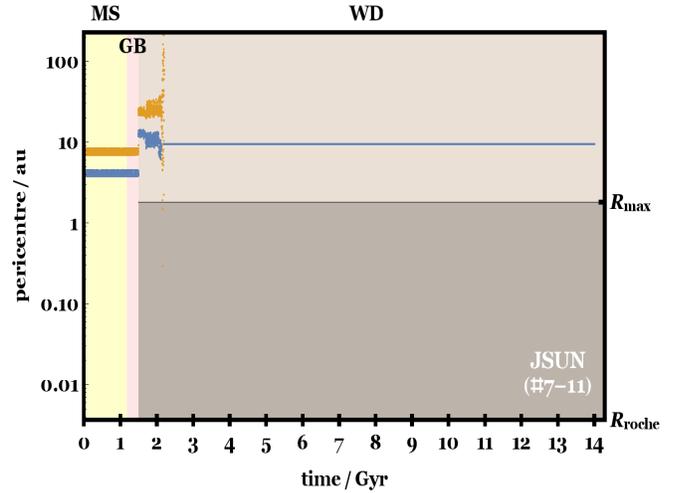,width=9cm,height=7cm}
\caption{
Evolution of a Solar-system analogue ({\tt JSUN}) that immediately ejects
Uranus and Neptune and keeps Saturn bound until the star is 2.2 Gyr old, which
is 0.7 Gyr into the white dwarf phase. Only Jupiter survives for this particular
evolution, which is simulation \#7-11 (Table \ref{JSUN}). 
}
\label{sym7-11}
\end{figure}

Next consider a Solar system analogue architecture ({\tt JSUN} and $\beta = 7$) from Table \ref{JSUN}.
Simulation \#7-11, shown in Fig. \ref{sym7-11}, features Neptune and Uranus being ejected at
10.3 Myr and 18.5 Myr (effectively immediately), which is not discernable on the plot. 
The remaining Jupiter and Saturn mutually perturb each other so that their pericentres vary
significantly (over 1 au in each case) throughout the main sequence. The orbital expansion 
causes the two-planet stability threshold (see \citealt*{debsig2002}, \citealt*{veretal2013a} 
and \citealt*{voyetal2013}) to be passed or at least skirted, leading to delayed instability
on the white dwarf phase.  The result is that at 2.2 Gyr, Saturn is ejected.  Jupiter
remains the lone survivor. 

\begin{figure}
\psfig{figure=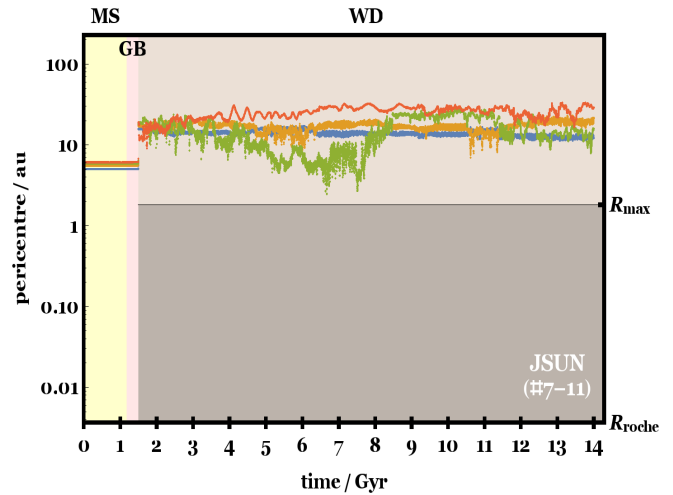,width=9cm,height=7cm}
\caption{
Unpacking of a set of terrestrial planets at the start of the white dwarf phase.
The planets ({\tt \=J\=S\=U\=N}) are scaled-down versions of Jupiter, Saturn, Uranus
and Neptune, with the mass reduced by a factor of about 318, which transforms Jupiter
into the Earth. All four planets remain stable, meander, and survive until the end of 
the simulation.  Shown is simulation \#7-40 (Table \ref{JSUN}). 
}
\label{sym7-40}
\end{figure}

\subsubsection{Solar system analogue -- terrestrial planets}

Alternatively, simulation \#7-40 ($\beta = 11$, and Fig. \ref{sym7-40}) illustrates one evolutionary
sequence for the scaled-down (by a mass factor of 318) versions of Jupiter, Saturn, Uranus and Neptune
({\tt \=J\=S\=U\=N}) -- effectively transforming them into terrestrial-mass planets.  
The system does not unpack until the EWD phase, but never 
becomes unstable.  The resulting meandering causes the scaled-down Uranus (green) to achieve 
an orbital pericentre of just 2.47 au (less than half of any planet's initial pericentre) at 6.67 Gyr.

\subsubsection{Terrestrial planet pericentre repacking}

\begin{figure*}
\centerline{
\psfig{figure=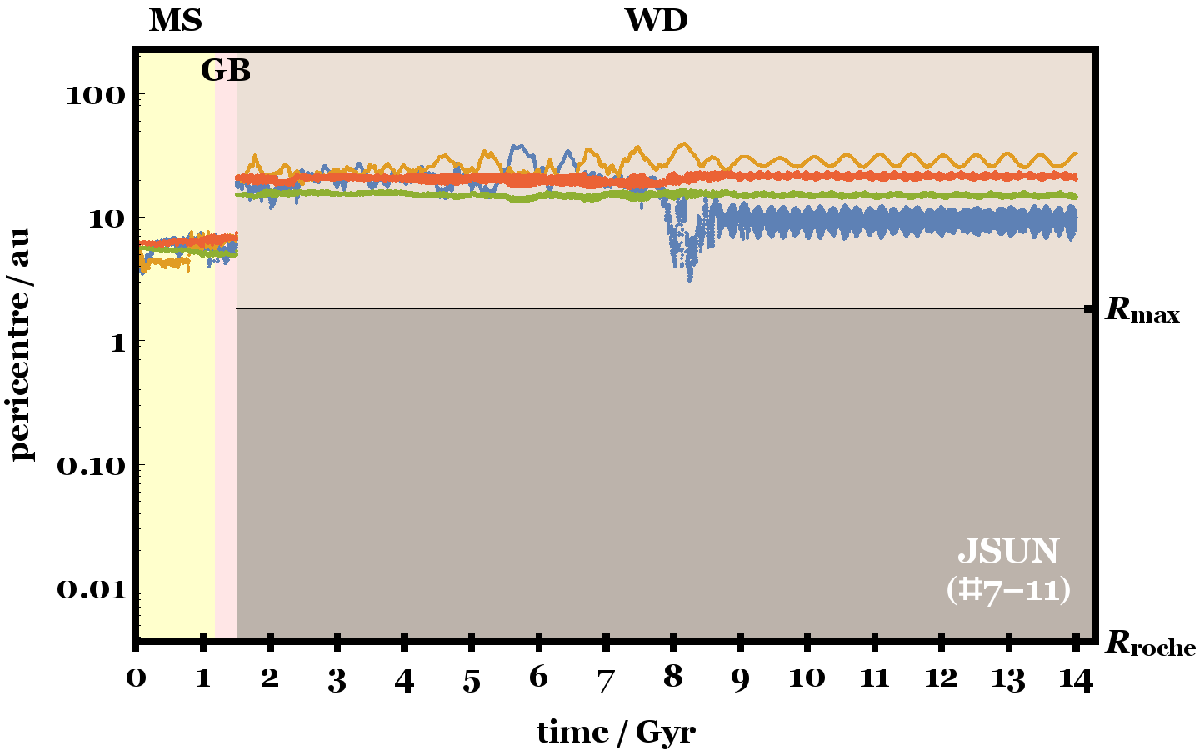,width=9cm,height=7cm}
\psfig{figure=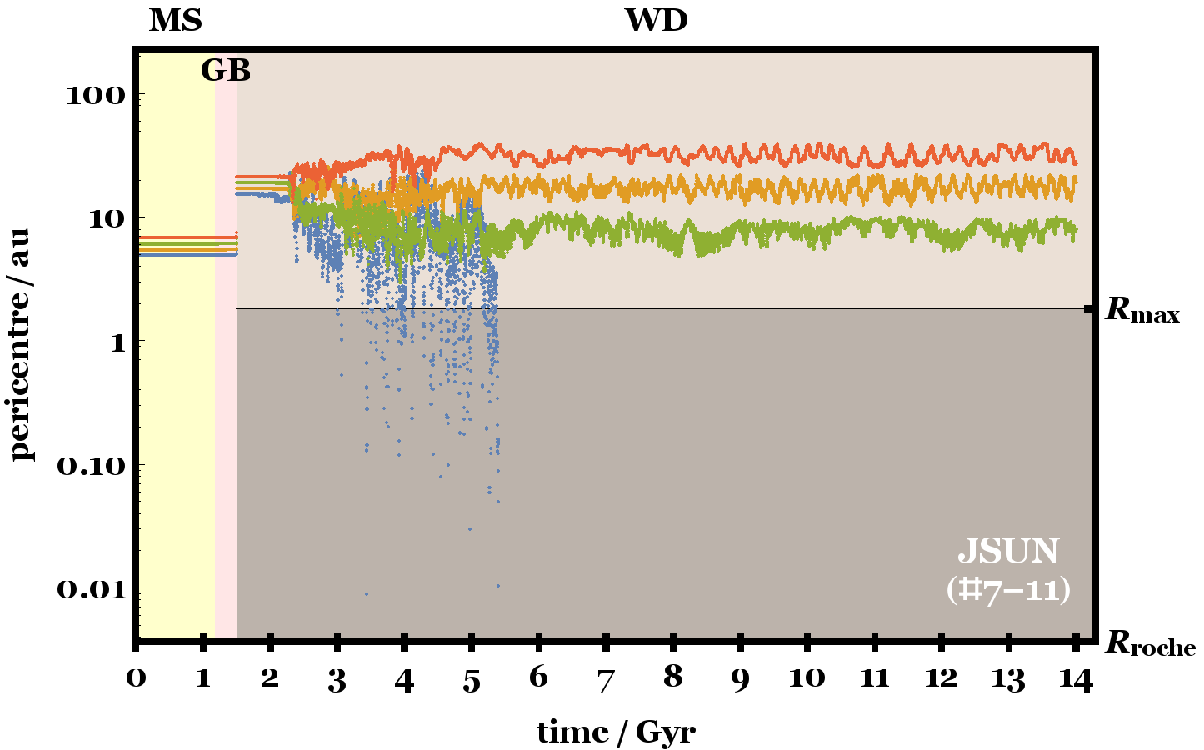,width=9cm,height=7cm}
}
\caption{
Two examples of unpacking then repacking, in both stable (left panel)
and unstable (right panel) cases. {\it Left Panel:}
Immediate unpacking of a set of terrestrial planets, followed by a re-ordering
and stable secular evolution from 8 Gyr onwards. The initial order of the planets
is {\tt \=U\=N\=J\=S} but their orbits end up in the order {\tt \=U\=J\=S\=N}.
Shown is simulation \#9-39 (Table \ref{UNJS}).
{\it Right Panel:} Unpacking of {\tt \=U\=J\=J\=J} on the WD phase, where the
least massive, innermost, planet becomes engulfed and two of the other planets
(orange and green) switch order. Shown is simulation \#2-24 (Table \ref{UJJJ}).
}
\label{sym9-39}
\end{figure*}

Another example of a long-term stable terrestrial system, but one that becomes unpacked immediately,
is from simulation \#9-39 ({\tt \=U\=N\=J\=S} -- blue, orange, green, red -- from Table \ref{UNJS} and left panel of
Fig. \ref{sym9-39}). This simulation contains two notable features: (i) the inward radial incursion of {\tt \=U} to a few au 
at around 8 Gyr (the first such radial incursion during the entire evolution), and (ii) the ``re-packing''
of the orbital pericentres beyond 8 Gyr. At this time, the system becomes 
orderly (but now in the order {\tt \=U\=J\=S\=N}) 
and henceforth secularly evolves with well-defined and periodic oscillations.

A second example of a repacked system, but one which becomes unstable, is illustrated in the right panel of Fig. \ref{sym9-39} ({\tt \=U\=J\=J\=J} from simulation \#2-24 of Table \ref{UJJJ}).  Here, the unpacking occurs on the LWD phase, the smallest-mass planet is engulfed, and the two closest {\tt \=J} planets switch places.

\subsubsection{Unpacked giant planets with test particles}

In reality, the above systems likely harbour Mars-like planets or asteroids in regions
like our Solar system's asteroid belt. Simulation \#11-19 ({\tt JUNS} and $\beta = 9$
from Table \ref{JUNS}) 
contains 12 test particles located in initially circular orbits at 2.5 au.  Figure \ref{sym11-19} shows
the resulting evolution. The four giant planets remain packed and stable through the entire
simulation, and have a nondisruptive effect on the test particles during the main sequence
and giant branch phases of evolution. However, on the white dwarf phase, eight of the 12 particles
are lost.  Seven are lost through ejections, all of which can be individually discerned on the
plot (at times 6.22, 7.49, 7.71, 8.20, 10.08, 10.66 and 12.19 Gyr). One particle is engulfed
inside of the white dwarf (as indicated in the table) at 10.29 Gyr.

\subsubsection{Deep radial incursions for almost equal-mass planets}

Tables \ref{JSJS} and \ref{UNUN} give details of simulations which contain almost equal-mass planets,
and therefore serve as a useful basis of comparison to both other simulations in this work
and previous simulations of strictly equal-mass planets.  The mass ratios of consecutive planet pairs 
in the tables are just 3.34 and 1.18, respectively.  

In Fig. \ref{symComp}, we display four simulations which show examples of how small ranges
in planet mass within the same system can lead to deep radial incursions during the white dwarf phase.
Shown are four-, six- and eight-planet systems.
In three of the cases (simulations \#12-8, \#12-21 and \#13-12) the runs did not finish.
Unpacking occurs on the WD phase in all cases, and instability results. The effects of tides
(not modelled) might affect the green planets (which achieve pericentres of $\lesssim 0.1$ au) 
on the bottom plots.

\subsection{The variety of system behaviours}  \label{sysbeh}

Having illustrated some specific examples, now we consider the simulations in aggregate.  
Before inferring trends from the data, we first consider the rich variety of behaviours and outcomes seen in the simulations
and simply list what is possible from full-lifetime evolution for clarity. 

\begin{itemize}
 \item Unpacking (defined as crossing orbits or planet loss) may occur during any phase of stellar evolution, or not at all.

 \item Unpacking through crossing orbits does not necessarily lead to instability (defined as planet loss from collisions or ejections)

 \item Unpacking during one phase can lead to instability at a later phase.

 \item Planet engulfment into the star, planet-planet collisions and ejections may all occur during any phase.

 \item Two systems with identical initial numbers, masses and separations of planets can be unpacked at different phases and lose different numbers of planets.

 \item Any total number of planets may be lost.

 \item Planets which are formed when the star arrives on the main sequence at distances well outside of the maximum asymptotic giant branch stellar radius can be perturbed on the white dwarf phase to distances well within the maximum asymptotic giant branch stellar radius.

 \item Test particles which initially reside within the orbits of four giant planets can survive for the
entire simulation duration even when the giant planets unpack and/or become unstable.
\end{itemize}

\subsection{General trends}   \label{gentre}

In this section we present the crux of our results and some trends
with applications beyond this work.

\subsubsection{Relating to $\beta$}
\begin{itemize}

\item Unpacking tends to occur at later stellar phases as $\beta$ is increased. This correlation is typically strong but by no means monotonic. For example, consider simulations \#8-1 through \#8-32 (Table \ref{JSNU}), where $\beta$ is increased from 6.0 to 9.5.  For a weaker correlation, instead see simulations \#9-1 through \#9-32 (Table \ref{UNJS}), and for a better correlation, see simulations \#4-1 through \#4-24 (Table \ref{UUUJ}).  See Fig. \ref{BetaPlot} for a visual representation of the correlation (although the tables themselves might be clearer).

\item Mapping a particlar value of $\beta$ to the phase at which one could expect unpacking is architecture-dependent.  Compare for example, the simulations with $\beta = 7.0$ across all of the tables.

\item Terrestrial-mass planets (effectively, {\tt \=J}, {\tt \=S}, {\tt \=U}, and {\tt \=N}) at a given $\beta$ will unpack at an earlier phase than their giant-planet counterparts (implied from Tables \ref{JUUU}-\ref{JUNS} and Fig. \ref{BetaPlot}) due to the additional dependence of stability timescale on mass, which is not captured by the Hill radius \citep[e.g.][]{chaetal1996,fabqui2007,musetal2014}.

\end{itemize}

\begin{figure}
\psfig{figure=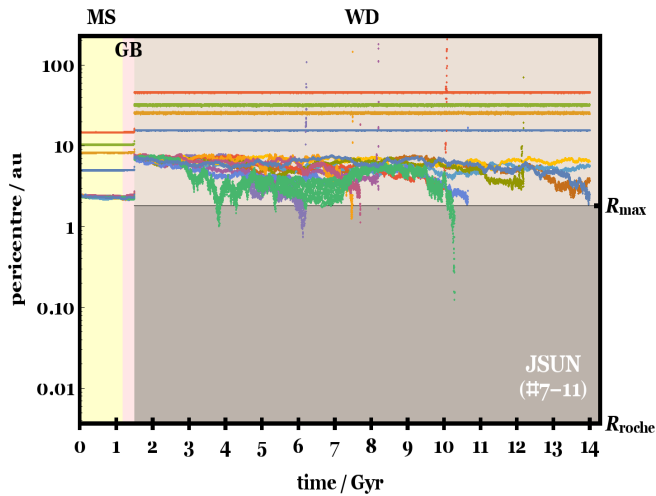,width=9cm,height=7cm}
\caption{
A stable and unpacked set of giant planets {\tt JUNS} with 12 test particles
initially located at 2.5 au. Four particles survive, seven are ejected, and one enters
the white dwarf disruption radius at 10.29 Gyr. All these events are visible on the plot.
Shown is simulation \#11-19 (Table \ref{JUNS}). 
}
\label{sym11-19}
\end{figure}
\begin{figure*}
\centerline{
\psfig{figure=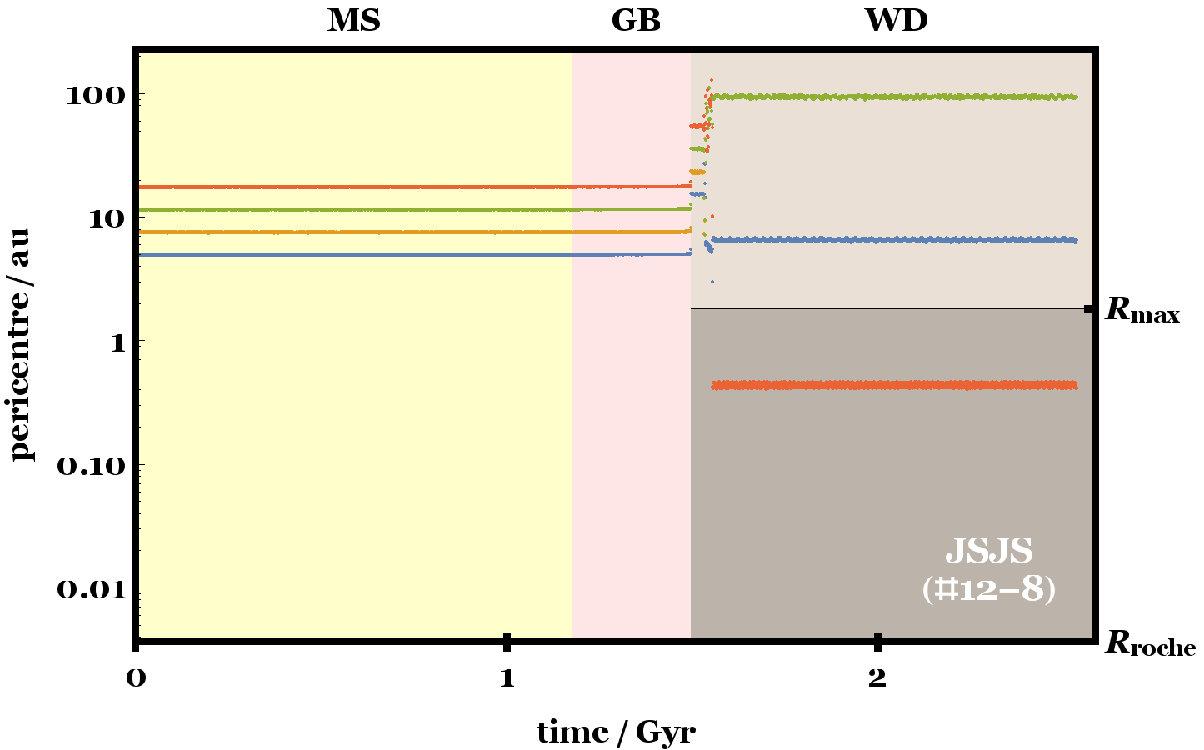,width=9cm,height=7cm}
\psfig{figure=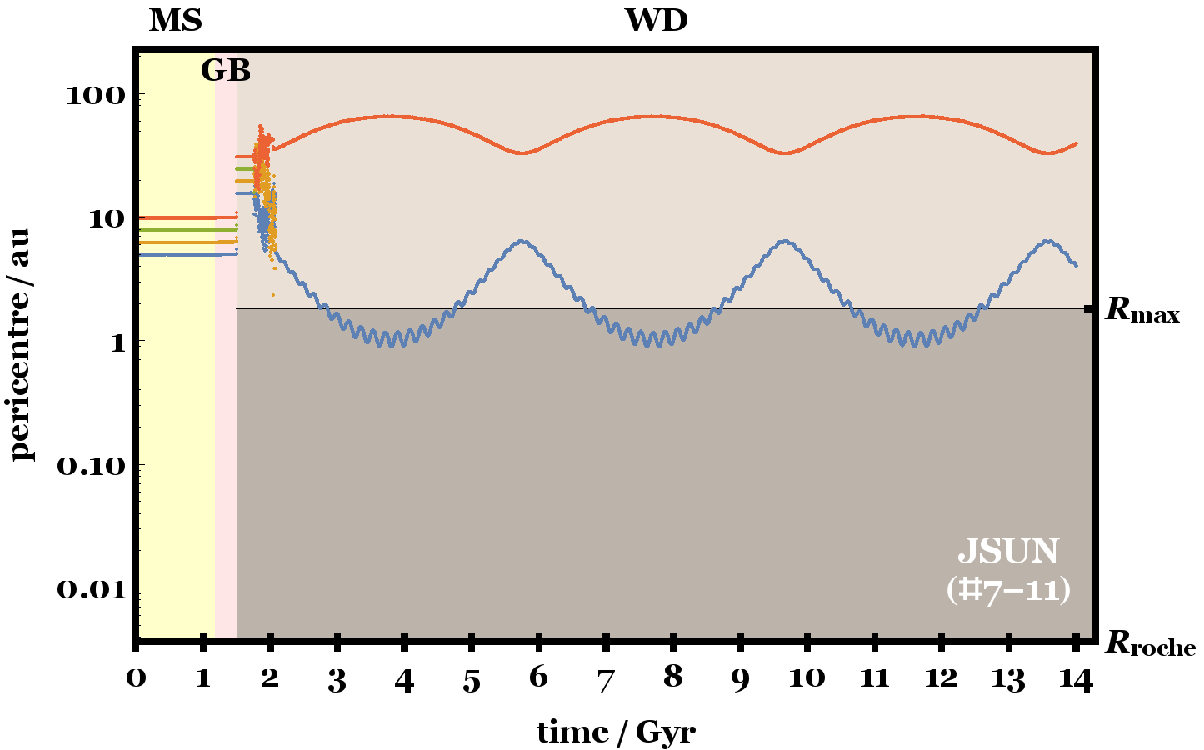,width=9cm,height=7cm}
}
\centerline{
\psfig{figure=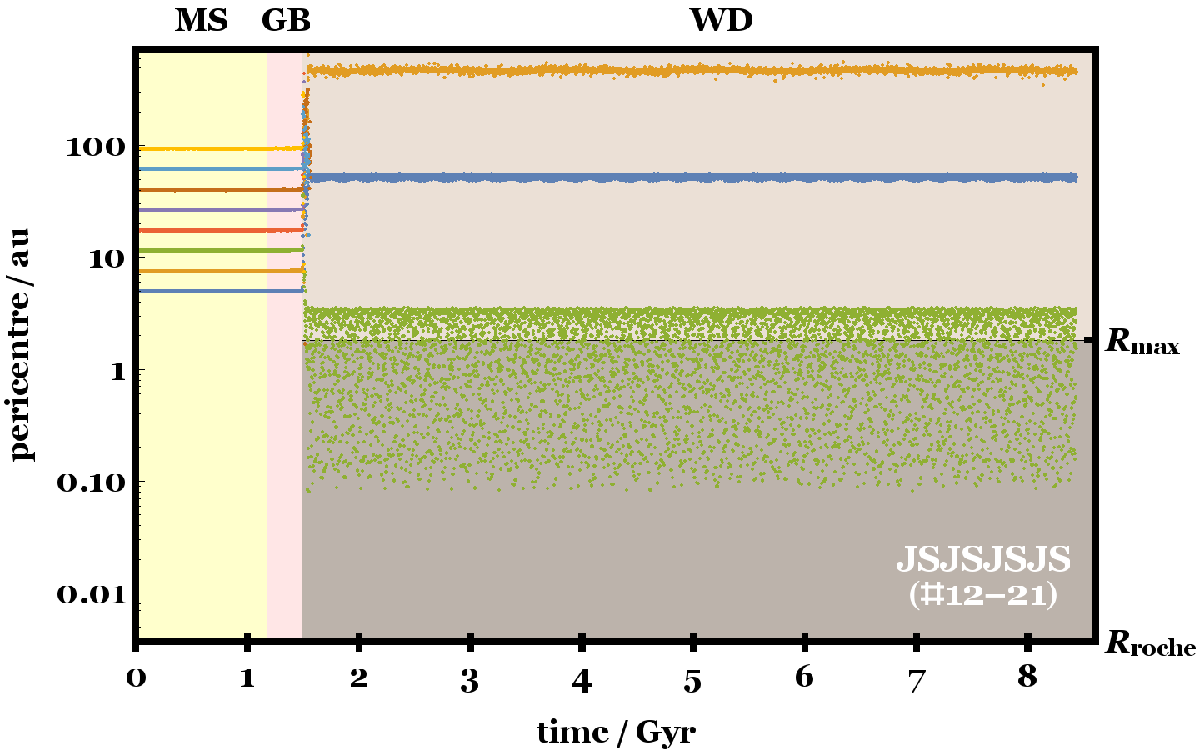,width=9cm,height=7cm}
\psfig{figure=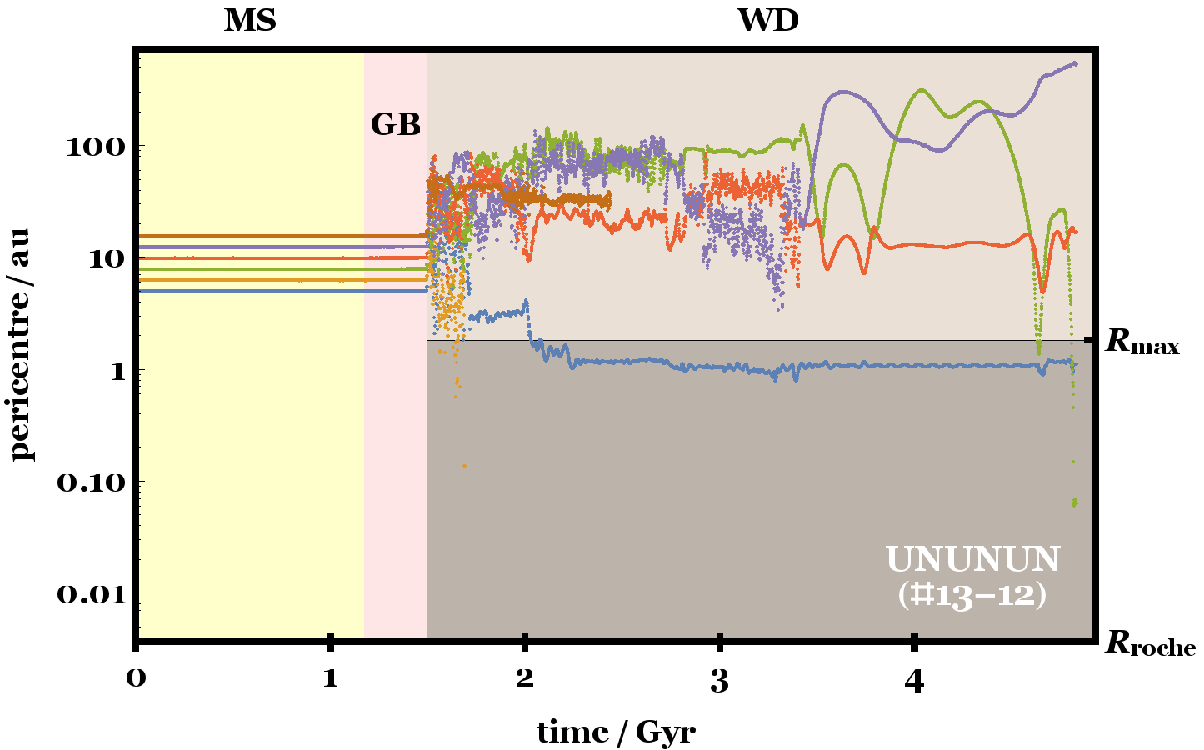,width=9cm,height=7cm}
}
\caption{
Deep radial incursions due to unpacking on the white dwarf phase of 
similar-mass giant planets in the systems {\tt JSJS} (upper left panel, simulation \#12-8),
{\tt JSJSJSJS} (lower left panel, simulation \#12-21), {\tt UNUN} (upper right panel, simulation \#13-5)
and {\tt UNUNUN} (lower right panel, simulation \#13-12). Tides are unlikely
to play a role in the upper panels, but might affect the evolution of the green planets
in the lower panels.  The striking behaviour seen here is characteristic of the simulations
from Tables \ref{JSJS} and \ref{UNUN}. 
}
\label{symComp}
\end{figure*}

\subsubsection{Relating to engulfments, ejections and collisions}
\begin{itemize}

\item Instability manifests itself primarily through ejections for giant planet systems 
and primarily through planet-planet collisions for terrestrial-planet systems. The two
stark exceptions are the architectures {\tt \=U\=J\=J\=J} and {\tt \=J\=J\=J\=U} (Tables \ref{UJJJ}
and \ref{JUUU}, where the lowest-mass terrestrial planet is engulfed into the white dwarf
in the majority of cases (see e.g. Fig. \ref{sym9-39}).

\item In-between these two regimes (giant planets and terrestrial planets) are the 
low-mass giant planets, or ice giants, with {\tt UNUN}, {\tt UNUNUN} and {\tt UNUNUNUN} 
(Table \ref{UNUN}). Only for these systems do unstable events appear to be roughly 
evenly distributed amongst ejections, engulfments and planet-planet
collisions. For simulations \#13-12 to \#13-17, the lack of planet-planet collisions might be due
to the truncated duration of those simulations and/or neglecting white dwarf-planet tides.

\item  Physically, the trends in the above two bullet points are understandable in terms of the 
Safronov number \citep{safzvj1969}, which is the square of the ratio of the surface escape speeds to the planetary orbital speeds. As this ratio increases, the frequency of 
ejections increases.  This ratio is approximately unity for Earth-like planets at 20 au,
but about 40 for Jupiters at the same separation.

\item The commonality of planet-planet collisions in terrestial planet systems implies
that those systems should contain more debris and newly-generated asteroids than giant
planet systems.

\item The unpacking of systems with four giant planets preferentially (80\%) results
in the survival of two planets. This percentage would be 90\% if not for the {\tt UNJS}
and {\tt NUJS} architectures (Tables \ref{UNJS} and \ref{NUJS}), which do not follow this trend.
In these architectures, either the Uranus or Neptune is typically ejected but the other
survives. 

\item The unpacking of systems with four terrestrial planets instead preferentially (55\%)
results in the survival of three planets, and in 30\% of cases retains all four planets.
This stark difference from the giant-planet case is likely related to the inability for
close encounters in terrestrial-planet systems to be strong enough to cause ejections.

\item Unpacked terrestrial-planet architectures which retain all planets are typically 
aperiodic in their resulting orbital variations (see e.g. Fig. \ref{sym7-40}). This feature is particularly 
noteworthy because these systems produce an ever-changing dynamic environment, which may 
tap into different reservoirs of white dwarf pollutants at different cooling ages.

\item When architectures contain one most massive planet (as opposed to two or more), as in 
Tables \ref{JUUU}, \ref{UUUJ}, \ref{JSUN}, \ref{JSNU}, \ref{UNJS}, \ref{NUJS} and \ref{JUNS},
that planet is {\it never} ejected nor engulfed into the star.  Physically, the reason is due
to conservation of angular momentum and energy, even though the system energy is strictly not conserved during GB mass loss.

\item For systems that contain exactly two most massive planets, those planets rarely
are ejected or engulfed into the star.  This tendency holds true for every single system 
simulated with Jupiters, Uranuses and their scaled equivalents {\tt JUUJ}, {\tt \=J\=U\=U\=J}, 
{\tt UJJU}, {\tt \=U\=J\=J\=U} (Tables \ref{JUUJ} and \ref{UJJU}).  For {\tt JSJS} (Table \ref{JSJS}), 
where the difference in planet mass is much less (ratio of 3 as opposed to 22), there is
only one exception (simulation \#12-6).  For {\tt UNUN} (Table \ref{UNUN}), there are two
exceptions.

\item Rarely (6.6\%) does unpacking of four-planet systems allow for at least one of the planets to 
eventually achieve an orbital pericentre within the maximum AGB radius of 1.8 au, in contrast to the equal planet-mass case
\cite{vergae2015}. 

\item Deep radial incursions are most common for the unequal-mass systems which are closest to the 
equal-mass case, namely the {\tt UJJJ}, {\tt \=U\=J\=J\=J},
{\tt JJJU}, {\tt \=J\=J\=J\=U}, {\tt JSJS} and {\tt UNUN} cases (Tables \ref{UJJJ}, \ref{JJJU}, \ref{JSJS} and \ref{UNUN}).
The reason for the similarity is in the first four cases when one ignores/ejects
the Uranus, and in the latter two cases because the range of their masses is small.  In that respect, the greatest
incidence of inward radial incursions occurs for the {\tt UNUN} architecture, because with a mass ratio
of 1.18 between adjacent pairs of planets, the system effectively contains equal-mass planets.

\item Increasing the number of planets in a system increases the incidence for deep inward radial incursions,
as well as consistently changing dynamical architectures, similar to the equal-planet mass case.

\end{itemize}

\begin{figure}
\psfig{figure=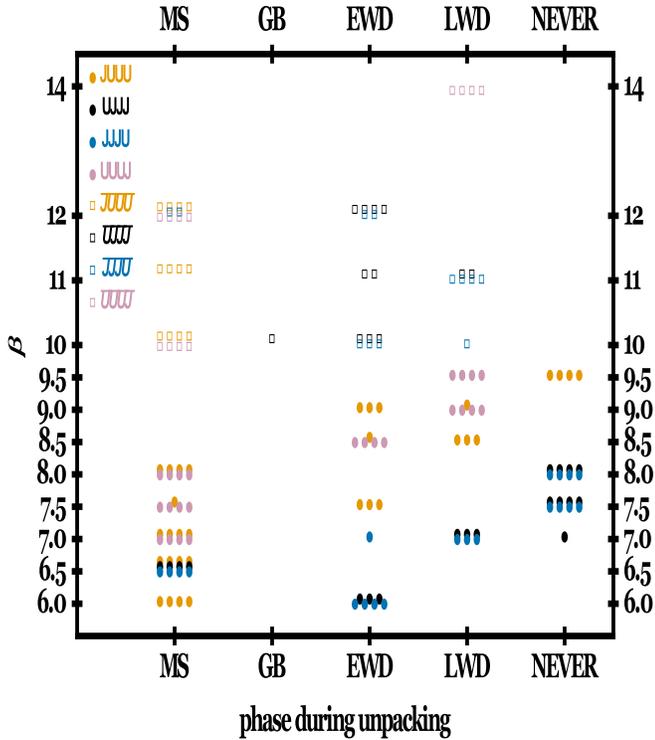,width=9cm,height=10.5cm}
\caption{
The phases at which unpacking occurs with respect to $\beta$ for the eight
architectures given in the legend. Each point represents one simulation.  
Generally, as $\beta$ is increased, unpacking occurs during later phases,
although the relationship is not monotonic. The terrestrial-sized planets
(open squares) generally require higher values of $\beta$ than giant planets
(dots) to achieve the same results.  Unpacking during the giant branch is rare.
}
\label{BetaPlot}
\end{figure}

\subsubsection{Relating to test particles}
\begin{itemize}

\item Unpacking of the non-zero mass planets enhances prospects for 
white dwarf engulfment of test particles, 
which can reasonably represent Mars-like planets or large asteroids.

\item Even with the tiny sample sizes adopted here (12 test particles per simulation, as constrained
by computational limitations), enough
are engulfed by white dwarfs (263 out of a total 1024) to suggest both that this process is crucial and 
that higher resolution studies are needed to detect discernable trends.

\end{itemize}

\section{Discussion}

In order to place our results in context, we will discuss the consequences for polluted white dwarf systems and consider the links to three outstanding observational constraints: (i) pollution rate with white dwarf cooling age, (ii) accumulated metal pollution in non-DA white dwarfs, and (iii) the WD\,1145+017 system.   We will also discuss the implications of so many ejections for the purported free-floating population of planets within the Milky Way, and how are simulations may be linked to chaos.

\subsection{Consequences for polluted white dwarf systems}

Our simulations clearly demonstrate that planet engulfment into white dwarfs is a rare phenomenon (8.8\% across
all simulations), in line with
the findings of the equal-planet mass studies of \cite{veretal2013a}, \cite{musetal2014} and \cite{vergae2015}.
A much more likely pollution reservoir is the test particles, which we have shown can easily be engulfed in
the white dwarf, in line with the one-planet studies
of \cite{bonetal2011}, \cite{debetal2012} and \cite{frehan2014}.  The difference here is that 
multiple planets provide the opportunity for a constantly changing dynamic environment, which
is not the case in one-planet systems\footnote{Stellar flybys can change the environment regardless
of the number of planets, but typically 10 Gyr needs to pass before a flyby achieves a close encounter
within a few hundred au \citep{zaktre2004,vermoe2012}.}. Consequently, multiple-planet systems are
much more likely to explain high rates of pollution at different cooling ages by accessing and
perturbing different reservoirs of material (asteroids, fragments, dust) at different times
and/or different locations.

Here we have characterized this environment by sampling systems of unequal-mass planets, where the planet
masses differ by a factor of up to about 20. We found that this inequality has clear but 
second-order effects on the dynamics; the first-order effects are determined by what types of planets
are involved in the unpacking: terrestrial or giant.  For giant planets, crossing orbits trigger violent encounters between 
giant planets, but still typically cause the system to settle into a periodic secular state (see Fig. \ref{sym1-19} and the upper-left panel of Fig. \ref{symComp}).
{\it For terrestrial planets, fully 30\% of our simulations become unpacked (orbit-crossed) but never 
unstable (featuring engulfments, ejections or collisions)}. The
result is a highly dynamic environment, where the planetary orbits {\it meander} (see Fig. \ref{sym7-40}), which is much more 
conducive to effective scattering at late ages.

\subsection{Correlation with cooling age}

\begin{figure*}
\centerline{
\psfig{figure=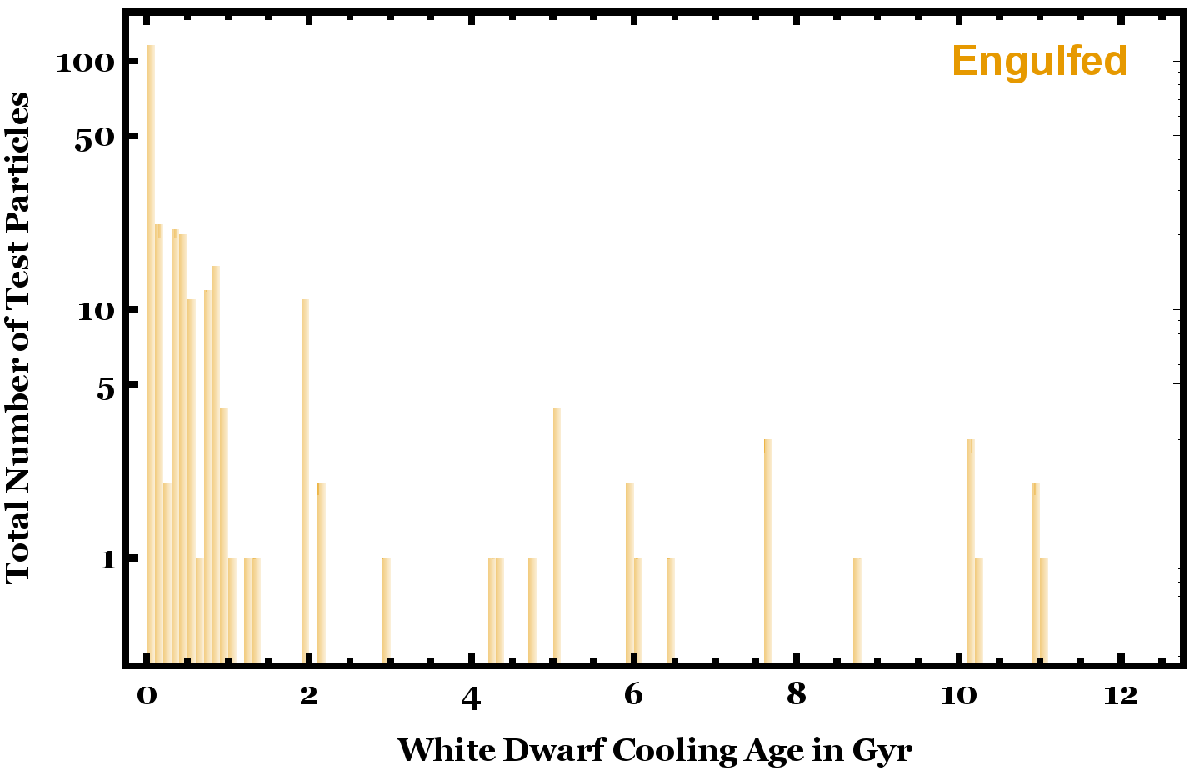,width=9cm,height=7cm}
\psfig{figure=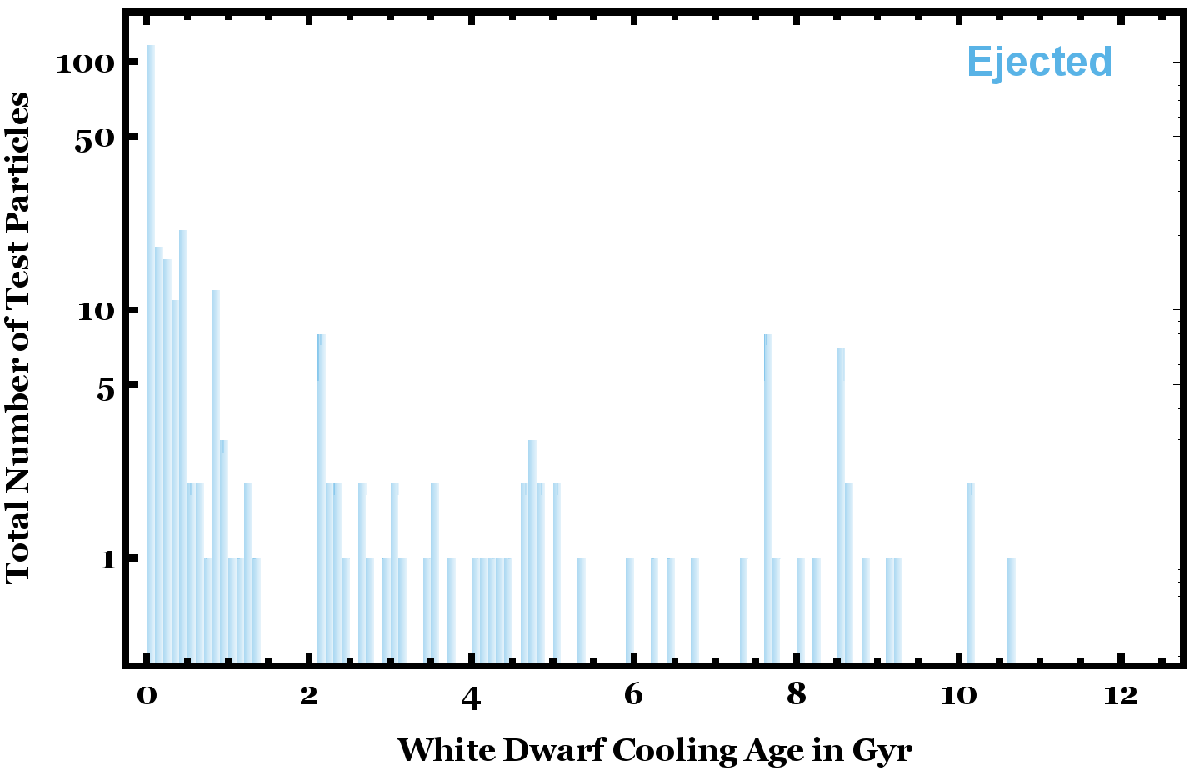,width=9cm,height=7cm}
}
\caption{
The white dwarf cooling age (time since becoming a white dwarf) at which test particles across
all simulations entered the white dwarf Roche radius ({\it left panel}) or were ejected from the system ({\it right panel}).  The EWD phase corresponds to the first bin, and the LWD phase to all other bins.  The histograms illustrate the pollution decay rate obtained from the simulations.
}
\label{CoolAge}
\end{figure*}

Our choice of dividing up the white dwarf phase into separate EWD and LWD phases was partly motivated by our simulation results, because a white dwarf cooling age of 100 Myr is a representative end value for the epoch of rapid post-main-sequence planetary instability (see e.g. Fig. \ref{CoolAge}). However, this value is also sensible from an observational point-of-view.  The cooling ages of the white dwarfs in \cite{koeetal2014} are all below 200 Myr, while white dwarf atmospheric properties can significantly change at cooling ages of $\sim 500$ Myr (see their Fig. 8, middle panel).  Therefore, a cut at cooling ages of a few 100 Myr is a natural way to separate samples observationally.

However, observations obtained so far indicate that the accretion rate of metals onto white dwarf atmospheres remains a flat function of white dwarf cooling age \citep[Fig. 4 of][]{koeetal2014}.  Explaining pollution at late times (after many Gyr of white dwarf evolution) is challenging because instability on the white dwarf phase is partially triggered by the increase in system stochasticity due to RGB and AGB mass loss \citep{voyetal2013}, preferentially leading to instabilities at early cooling ages. Fig. \ref{CoolAge} emphasizes this tendency, even though this study does not attempt to model a realistic population synthesis (which is anyway beyond current computational means). 

Recent and ongoing work is exploring potential ways of polluting white dwarfs at late cooling ages.  One possibility is through the change in orbits of wide binary stellar companions due to Galactic tides after, and only after, one of the components has become a white dwarf \citep{bonver2015}. However, the majority of known polluted white dwarfs do not appear to harbour wide-orbit companions.  Another possibility is through Lidov-Kozai secular evolution amongst multiple planets, such that the close encounters between planets and white dwarfs first occur only after cooling ages of several Gyr (C. Petrovich \& D. Mu\~noz, in preparation). Finally, an extant fragment field from planet-planet collisions may persist for several Gyr before being thrust towards the white dwarf (A. Shannon et al., in preparation).

For the architectures we have explored here, there is a similar spike in instabilities just after mass loss from the tip of the AGB during the EWS phase. However, planetary systems which remain stable through that epoch exhibit a wide range of instability times, and instances when a planet or test particle approaches the vicinity of the white dwarf.  Meandering of low-mass (terrestrial-like) planets provides a dynamic environment with which extant debris or fragments may be perturbed to the white dwarfs at all ages. Our results show that mass equality amongst planets is not a requirement for late-age pollution, and is not in fact even preferential for producing instabilities at late ages.

\subsection{Accumulated metals in convection zone}

White dwarfs with deep convection zones (usually containing Helium-dominated atmospheres) retain a measurable record of the accreted planetary debris over a span of time up to a few Myr (see Fig. 1 of \citealt*{wyaetal2014}).  Fig. 6 of \cite{veras2016} illustrates the amount of mass accreted for three different samples from \cite{faretal2010}, \cite{giretal2012} and \cite{xujur2012}.

The accumulated mass ranges from the mass of Phobos to that of Pluto, and may have been accrued by a single object or a collection of bodies. Distinguishing these two possibilities is not possible observationally. From theory, we may determine the likelihood of a sequence of bodies impacting the white dwarf or entering its Roche radius within ~1 Myr.  However, the sample size of the test particles in our simulations here (12 per simulation) was too small to determine impact frequency for a given architecture. 

The accretion itself might represent a combination of a ``continuous'' stream of small particles from a surrounding disc and a ``stochastic'' agglomeration of larger particles from elsewhere in the system. \cite{wyaetal2014} showed that the size boundary between these two regimes is approximately 35 km, and further constrained the potential size distribution of this accreted material, ruling out a mono-mass distribution.  Further, discs have been detected around only a few percent of polluted white dwarfs \citep{faretal2009,giretal2011,steetal2011}, although the actual fraction is likely greater than half \citep{beretal2014}.  Consequently, stochastic accretion is likely to play a role in many of these systems. The mechanics of impact into white dwarf atmospheres indicates that the parameter space may be split into sublimation, fragmentation and ablation regimes (J.~C. Brown et al., in preparation) such that the details of the deposition are complex, but the end result is still metals in the convection zone.

\subsection{WD 1145+017}

The WD 1145+017 planetary system represents the only example of a metal-polluted white dwarf with a surrounding debris disc composed of both dust and gas and disintegrating objects (asteroids, planets, or something in-between) detected by transit photometry. In this respect, the system provides a self-consistent snapshot of the disc formation and accretion process that is likely to take place at other metal-polluted white dwarfs, and confirms long-standing theories \citep{graetal1990,jura2003,beasok2013}.

The system was announced by \cite{vanetal2015}, who presented transit curves that illustrated that up to six objects with orbital periods of about 4.5-4.9 hours are in the process of disintegrating and producing dust. They found that the dominant orbital period is closer to 4.5 hours, which places the objects near the white dwarf's Roche radius, assuming that the objects are rubble-piles like the asteroids seen in the Solar system.  The size of these objects are poorly constrained, and could range anywhere from $\sim 1$ to $1000$ km. We will henceforth refer to them as planetesimals.

Follow-up observations came quickly \citep{croetal2016,gaeetal2016,rapetal2016,xuetal2016}. \cite{xuetal2016} detected gas in the debris disc and showed that the white dwarf atmosphere is polluted with 11 heavy elements.  \cite{croetal2016} performed multi-wavelength observations that illustrated the number of planetesimals disintegrating is likely more than one, and helped confirm that the planetesimals harbour an orbital period of about 4.5 hours rather than a value closer to 4.9 hours.  Further follow-up was provided by \cite{gaeetal2016}, who used high-speed photometry and observations of the system from Nov-Dec 2015 to reveal that at least six planetesimals are breaking up, and that they share the same near-circular orbit with orbital periods of about 4.4930 hours.  \cite{rapetal2016} most recently detected drifting features which they postulate are fragments that broke off from a single progenitor.

Our results, along with those of \cite{bonetal2011}, \cite{debetal2012}, \cite{frehan2014} and \cite{vergae2015}, demonstrate that the progenitor of the planetesimals in WD 1145+017 may be a large asteroid that was scattered in the vicinity of the WD.  The scattering may be caused by one planet \citep{bonetal2011,debetal2012,frehan2014} or multiple planets (this paper).  Alternatively, the progenitor may be a moon \citep{payetal2016}, or a small (terrestrial) planet, as shown by both \cite{vergae2015} and this paper.  Multiple planets can scatter a test particle into a transit-detectable orbit, even if the planets themselves never unpack (Fig. \ref{sym11-19}).  We note that a Solar system analogue, with {\sc JSUN} and asteroids or a Mars, can easily generate the progenitor of the planetesimals in WD 1145+017.  The mass of the progenitor remains unconstrained \citep{veretal2016}.

Our simulations suggest that WD 1145+017 is not unique in hosting transiting planetesimals.  Many possible multi-planet scenarios can perturb test particles into the Roche radius of the white dwarf; we have just scratched the surface.

\subsection{Free-floating planet population contribution}

A brief inspection of the tables in the appendix reveal a preponderance of planetary ejections.  This feature, just like for the equal-mass planet case \citep{veretal2013a,musetal2014}, and even the single-planet case \citep{veretal2011,vertou2012,veretal2014c},  help establish that planetary ejection is an ubiquitous feature of post-main-sequence systems. Consequently, these ejections make a contribution to the free-floating planet population.

How this contribution compares to that due to the dynamical activity which accompanies planetary formation and protoplanetary disc dissipation \citep[e.g.][]{rasfor1996,levetal1998,marwei2002,verarm2005,verarm2006, rayetal2011,rayetal2012,matetal2013} is not yet clear, primarily because our observational knowledge of exoplanets beyond 5 au is sparse. Nevertheless, we have an extraordinary observational estimate on the total number of free-floating giant planets in the Milky Way: nearly two for every main sequence star \citep{sumetal2011}. If future observations affirm this result, then a major question in planetary science will remain identifying the origin of so many free-floaters.  Planetary scattering alone with single stars on the main sequence cannot explain this population \citep{verray2012}, and the contribution from scattering in binary systems has not yet been quantified, despite studies such as \cite{sutfab2015} and \cite{smuetal2016}.

We caution that although ejections were perhaps common amongst the currently observed population of white dwarfs (as illustrated by this study), the resulting contribution to the currently observed free-floating planet population as reported by \cite{sumetal2011} would be on the order of 1\% \citep{verray2012}.  Each white dwarf progenitor system would need to have harboured tens of {\it giant} planets to achieve the \cite{sumetal2011} result.  This number is thought to be too large, despite the likely positive correlation between stellar mass and planet multiplicity \citep{kenken2008,andetal2013}, partly because of the extreme case HR 8799, which contains (just) four giant and packed planets \citep{maretal2010}.  HR 8799 also provides a representative glimpse into the past of white dwarf planetary systems because of its A-star host.

\subsection{Linking meandering with chaotic behaviour}

The phenomenon we refer to as meandering is linked to the stochasticity of the system.  The vast literature on chaos indicators in gravitational point-mass exoplanetary systems utilizes a wide variety of techniques in order to characterize, in part, how close the system is to instability at different times.  Linking these indicators to $N$-body integrations -- particularly long-term integrations -- remains challenging (e.g. Veras, Antoniadou \& G\"{a}nsicke, In Prep) but may provide key constraints.

Only two published dedicated post-main-sequence studies of which we are aware have attempted to link evolution with stochasticity at the beginning and end of mass loss \citep{adaetal2013,voyetal2013}.  Both studies use the classical Lyapunov exponent as their chaos indicators, and consider two planets.  Their work brings attention to subtleties which indicate that dedicated studies on chaos would be beneficial.  We frame these subtleties as the following questions: (i) What metric (e.g. Cartesian coordinate, eccentricity), and corresponding reference frame or coordinate system, would provide the most representative link to instability?, (ii) How does one combine chaos indicators for multi-planet systems, particularly after close encounters, and after one or more of the planets is lost from the system?, (iii) What chaos indicators are affected by the Hamiltonian-breaking physics of stellar mass loss, and how do different mass loss prescriptions affect the usefulness of a particular indicator?

Although these questions are too big to tackle here, our simulations provide a template on which future dedicated studies may make useful comparisons.

\section{Conclusions}

We have performed over 450 full-lifetime simulations of unequal-mass planets, which finally removes the long-standing equal-mass constraint from previous studies. We have also for the first time simulated the post-main-sequence evolution of multiple planets with test particles. Appendix A displays the results and characteristics of all simulations. The trends in the data are outlined in bulleted form in Section \ref{gentre}, and are summarized here as

\begin{itemize}

\item Unlike in the giant planet case, terrestrial-planet unpacking (orbit crossing) often does not trigger instability (engulfments, ejections and collisions), and provides a more dynamic, constantly shifting evolution throughout the white dwarf phase; this result is independent of the mass variation amongst planets.

\item The smaller the dispersion in planetary mass, the closer those planets may be perturbed towards the white dwarf.

\item Giant planet systems preferentially feature ejections whereas terrestrial-planet systems preferentially feature planet-planet collisions. Consequently, we expect more potentially polluting debris to exist in terrestrial-planet systems.

\item Prospects for unpacking roughly increase as $\beta$ increases, although this relationship is not-monotonic and dependent on the considered architecture.

\end{itemize}

Ultimately, how planets behave at different phases of evolution will crucially determine the subsequent evolution of the smaller bodies in those systems, bodies which are most likely the progenitors of white dwarf pollution and planetesimals such as those observed disintegrating around WD 1145+017.

\section*{Acknowledgements}

DV and BTG have received funding from the European Research Council under the European Union's Seventh Framework
Programme (FP/2007-2013)/ERC Grant Agreement n. 320964 (WDTracer). AJM acknowledges support from grant number
KAW 2012.0150 from the Knut and Alice Wallenberg foundation and the Swedish Research Council (grant 2011-3991).

\appendix

\section{Simulation Data Tables}

In this appendix, we present characteristics of every simulation, one per row. The simulation sets are split
into tables according to the masses and ordering of planets simulated.  See Sec. \ref{tabcol} for a full description of the table columns.




\begin{table*}
 \centering
 \begin{minipage}{180mm}
 \centering
  \caption{Summary of results for {\tt JUUU} and {\tt \=J\=U\=U\=U}.  We summarize the column definitions 
          (see Sec. \ref{tabcol} for a full description) as:  
          {\bf Sim \#}: Simulation designation.  
          {\bf Setup}: Planet type and order from closest to furthest. Overbars denote a mass reduction by a factor of 318.
          {\bf $\boldsymbol{\beta}$}: Number of mutual Hill radii. 
          {\bf Unpack}: Stellar phase during which unpacking occurs.
          {\bf \# Surv}: Number of surviving planets
          {\bf Engulf}: Planets (identified in subscripts in number order from closest to furthest) which intersect the star's surface or Roche radius, and the phase when the engulfment occurs.
          {\bf Eject}: Planets (identified in subscripts in number order from closest to furthest) which are ejected from the system, and the phase when the ejection occurs.
          {\bf Collision}: Planets (identified in subscripts in number order from closest to furthest) which collide with one another, and the phase when the collision occurs.
          {\bf $\boldsymbol{<} \boldsymbol{R_{\rm max}}$}: Surviving planets (identified in number order from closest to furthest) which acheive an orbital pericentre less than 1.82 au during the EWD or LWD phase; the minimum pericentre is provided in the subscript.
          {\bf TPs Eng}: Number of test particles out of 12 which are engulfed in the EWD/LWD phases.  
          {\bf Phase abbreviations}:
          MS = main sequence, GB = giant branch, EWD = white dwarf with 0-100 Myr cooling, LWD = white dwarf beyond 100 Myr cooling.\newline
          Notes:\newline
$^{\rm d}$A test particle which survived the entire integration acheived a minimum pericentre of 0.062 au at a WD cooling age of 1.983 Gyr.\newline 
                                                         }
          \label{JUUU}
  \begin{tabular}{@{}ccccccccccc@{}}
 \hline
Sim \#  &  Setup      & $\beta$ & Unpack    & \# Surv & Engulf & Eject           & Collision     & $< R_{\rm max}$? &  TPs Eng & Notes   \\
 \hline
1-1  &  {\tt JUUU}          &  6.0       & MS       & 2        &          & MS$_2$,MS$_3$     &                &         &  &  \\
1-2  &  {\tt JUUU}          &  6.0       & MS       & 2        &          & MS$_3$            &  MS$_{1-2}$     &         &  &  \\
1-3  &  {\tt JUUU}          &  6.0       & MS       & 2        &          & MS$_2$,MS$_3$     &                &         &  & \\
1-4  &  {\tt JUUU}          &  6.0       & MS       & 2        &          & MS$_2$,MS$_3$     &                &         &  & \\
\hline
1-5  &  {\tt JUUU}          &  6.5       & MS       & 2        &          & MS$_3$,MS$_4$     &                &         &  & \\
1-6  &  {\tt JUUU}          &  6.5       & MS       & 2        &          & MS$_3$,MS$_4$     &                &         &  & \\
1-7  &  {\tt JUUU}          &  6.5       & MS       & 2        &          & MS$_3$            &  MS$_{1-4}$     &         &  &  \\
1-8  &  {\tt JUUU}          &  6.5       & MS       & 2        &          & MS$_2$,MS$_3$     &                &         &  & \\
\hline
1-9  &  {\tt JUUU}          &  7.0       & MS       & 2        &          & MS$_2$,MS$_3$     &                &         &  & \\
1-10 &  {\tt JUUU}          &  7.0       & MS       & 2        &          & MS$_2$,MS$_3$     &                &         &  & \\
1-11 &  {\tt JUUU}          &  7.0       & MS       & 2        &          & MS$_3$,MS$_4$     &                &         &  & \\
1-12 &  {\tt JUUU}          &  7.0       & MS       & 2        &          & MS$_2$            &  MS$_{3-4}$     &         &  &  \\
\hline
1-13 &  {\tt JUUU}          &  7.5       & MS       & 2        &          & MS$_2$,MS$_4$     &                &         &  & \\
1-14 &  {\tt JUUU}          &  7.5       & EWD        & 2        &          & EWD$_3$,EWD$_4$       &                &         &  & \\
1-15 &  {\tt JUUU}          &  7.5       & EWD        & 2        &          & EWD$_3$,EWD$_4$       &                &         &  & \\
1-16 &  {\tt JUUU}          &  7.5       & EWD        & 2        &          & EWD$_2$,EWD$_4$       &                &         &  & \\
\hline
1-17 &  {\tt JUUU}          &  8.0       & MS       & 2        &          & MS$_3$,EWD$_4$      &                &         &  & d \\
1-18 &  {\tt JUUU}          &  8.0       & MS       & 2        &          & MS$_3$,MS$_4$     &                &         &  & \\
1-19 &  {\tt JUUU}          &  8.0       & MS       & 2        &          & MS$_2$,MS$_3$     &                &         &  & \\
1-20 &  {\tt JUUU}          &  8.0       & MS       & 2        &          & MS$_3$,MS$_4$     &                &         &  & \\
\hline
1-21 &  {\tt JUUU}          &  8.5       & EWD        & 2        &          & EWD$_3$,EWD$_4$       &                &         & (0/0) & \\
1-22 &  {\tt JUUU}          &  8.5       & LWD       & 2        &          & LWD$_2$,LWD$_4$     &                &         & (0/0) & \\
1-23 &  {\tt JUUU}          &  8.5       & LWD       & 2        &          & LWD$_2$,LWD$_3$     &                &         & (0/4) & \\
1-24 &  {\tt JUUU}          &  8.5       & LWD       & 2        &          & LWD$_2$,LWD$_4$     &                &         & (0/0) & \\
\hline
1-25 &  {\tt JUUU}          &  9.0       & LWD       & 2        &          & LWD$_2$,LWD$_3$     &                &         & (0/0) & \\
1-26 &  {\tt JUUU}          &  9.0       & EWD        & 2        &          & EWD$_3$,EWD$_4$       &                &         & (0/0) & \\
1-27 &  {\tt JUUU}          &  9.0       & EWD        & 2        &          & EWD$_3$,EWD$_4$       &                &         & (3/2) & \\
1-28 &  {\tt JUUU}          &  9.0       & EWD        & 2        &          & EWD$_3$,EWD$_4$       &                &         & (2/0) & \\
\hline
1-29 &  {\tt JUUU}          &  9.5       &          & 4        &          &                  &                &         & (0/0) &  \\
1-30 &  {\tt JUUU}          &  9.5       &          & 4        &          &                  &                &         & (0/0) &  \\
1-31 &  {\tt JUUU}          &  9.5       &          & 4        &          &                  &                &         & (0/0) & \\
1-32 &  {\tt JUUU}          &  9.5       &          & 4        &          &                  &                &         & (0/0) & \\
\hline
1-33 &  {\tt \=J\=U\=U\=U}  &  10.0      & MS       & 3        &          &                  &  MS$_{1-4}$     &         &  & \\
1-34 &  {\tt \=J\=U\=U\=U}  &  10.0      & MS       & 2        &          &                  &  MS$_{2-3}$,MS$_{1-2}$  &  &  &  \\
1-35 &  {\tt \=J\=U\=U\=U}  &  10.0      & MS       & 3        &          &                  &  MS$_{1-2}$     &         &  & \\
1-36 &  {\tt \=J\=U\=U\=U}  &  10.0      & MS       & 2        &          &                  &  MS$_{1-4}$,MS$_{1-2}$  &  &  &  \\
\hline
1-37 &  {\tt \=J\=U\=U\=U}  &  11.0      & MS       & 3        &          &                  &  MS$_{3-4}$     &         &  & \\
1-38 &  {\tt \=J\=U\=U\=U}  &  11.0      & MS       & 2        &          &                  &  MS$_{1-2}$,MS$_{3-4}$  &  &  & \\
1-39 &  {\tt \=J\=U\=U\=U}  &  11.0      & MS       & 2        &          &                  &  MS$_{1-3}$,LWD$_{1-2}$  &  &  & \\
1-40 &  {\tt \=J\=U\=U\=U}  &  11.0      & MS       & 4        &          &                  &                &         &  & \\
\hline
1-41 &  {\tt \=J\=U\=U\=U}  &  12.0      & MS       & 2        &          &                  &  MS$_{1-3}$,LWD$_{1-4}$  &  &  & \\
1-42 &  {\tt \=J\=U\=U\=U}  &  12.0      & MS       & 4        &          &                  &                &         &  & \\
1-43 &  {\tt \=J\=U\=U\=U}  &  12.0      & MS       & 3        &          & LWD$_3$           &                &         &  & \\
1-44 &  {\tt \=J\=U\=U\=U}  &  12.0      & MS       & 3        &          &                  &  MS$_{1-2}$     &         &  &  \\
 \hline
\end{tabular}
\end{minipage}
\end{table*}

\begin{table*}
 \centering
 \begin{minipage}{180mm}
 \centering
  \caption{Summary of results for {\tt UJJJ} and {\tt \=U\=J\=J\=J}.  See Sec. \ref{tabcol} for a full description of the columns or Table 
          \ref{JUUU} for a summary. \newline
          MS = main sequence, GB = giant branch, EWD = white dwarf with 0-100 Myr cooling, LWD = white dwarf beyond 100 Myr cooling.\newline
          Notes:\newline $^{\rm a}$Subsequent evolution may have been affected by tides on the GB phase.}
          \label{UJJJ}
  \begin{tabular}{@{}ccccccccccc@{}}
 \hline
Sim \#  &  Setup      & $\beta$ & Unpack    & \# Surv & Engulf & Eject        & Collision     & $< R_{\rm max}$? &  TPs Eng & Notes   \\
 \hline
2-1  &  {\tt UJJJ}          &  6.0       & EWD       & 2        &          & EWD$_3$,EWD$_4$     &                &    \#2$_{1.449}$     &  &  \\
2-2  &  {\tt UJJJ}          &  6.0       & EWD       & 2        & EWD$_1$    & LWD$_2$          &                &    \#4$_{0.448}$     &  &  \\
2-3  &  {\tt UJJJ}          &  6.0       & EWD       & 2        &          & EWD$_1$,EWD$_2$     &                &                     &  &  \\
\hline
2-4  &  {\tt UJJJ}          &  6.5       & MS      & 2        & MS$_1$   & MS$_2$          &                &                     &  &  \\
2-5  &  {\tt UJJJ}          &  6.5       & MS      & 2        &          & MS$_1$,EWD$_4$    &                &    \#3$_{1.114}$     &  &  \\
2-6  &  {\tt UJJJ}          &  6.5       & MS      & 2        &          & MS$_1$,LWD$_2$   &                &                     &  &  \\
2-7  &  {\tt UJJJ}          &  6.5       & MS      & 2        &          & MS$_1$          & MS$_{3-4}$     &                     &  &  \\
\hline
2-8  &  {\tt UJJJ}          &  7.0       & LWD      & 2        &          & LWD$_1$,LWD$_3$   &               &                     &  &  \\
2-9  &  {\tt UJJJ}          &  7.0       &         & 4        &          &                 &               &                     &  &  \\
2-10 &  {\tt UJJJ}          &  7.0       & LWD      & 2        &          & LWD$_1$,LWD$_3$   &               &                     &  &  \\
2-11 &  {\tt UJJJ}          &  7.0       & LWD      & 3        &          & LWD$_1$          &               &                     &  &  \\
\hline
2-12 &  {\tt UJJJ}          &  7.5       &         & 4        &          &                 &               &                     &  &  \\
2-13 &  {\tt UJJJ}          &  7.5       &         & 4        &          &                 &               &                     &  &  \\
2-14 &  {\tt UJJJ}          &  7.5       &         & 4        &          &                 &               &                     &  &  \\
2-15 &  {\tt UJJJ}          &  7.5       &         & 4        &          &                 &               &                     &  &  \\
\hline
2-16 &  {\tt UJJJ}          &  8.0       &         & 4        &          &                 &               &                     &  &  \\
2-17 &  {\tt UJJJ}          &  8.0       &         & 4        &          &                 &               &                     &  &  \\
2-18 &  {\tt UJJJ}          &  8.0       &         & 4        &          &                 &               &                     &  &  \\
2-19 &  {\tt UJJJ}          &  8.0       &         & 4        &          &                 &               &                     &  &  \\
\hline
2-20 &  {\tt \=U\=J\=J\=J}  &  10.0      & EWD       & 3        &          &                  &  EWD$_{1-3}$    &    \#2$_{0.833}$     &  &  \\
2-21 &  {\tt \=U\=J\=J\=J}  &  10.0      & EWD       & 3        & LWD$_1$   &                  &              &                     &  &  \\
2-22 &  {\tt \=U\=J\=J\=J}  &  10.0      & EWD       & 3        & LWD$_1$   &                  &              &                     &  &  \\
2-23 &  {\tt \=U\=J\=J\=J}  &  10.0      & GB      & 3        & LWD$_1$   &                  &              &                     &  & a \\
\hline
2-24 &  {\tt \=U\=J\=J\=J}  &  11.0      & LWD      & 3        & LWD$_1$   &                  &              &                     &  &  \\
2-25 &  {\tt \=U\=J\=J\=J}  &  11.0      & LWD      & 4        &          &                  &              &    \#4$_{0.989}$     &  &  \\
2-26 &  {\tt \=U\=J\=J\=J}  &  11.0      & EWD       & 3        & LWD$_1$   &                  &              &                     &  &  \\
2-27 &  {\tt \=U\=J\=J\=J}  &  11.0      & EWD       & 3        & LWD$_1$   &                  &              &                     &  &  \\
\hline
2-28 &  {\tt \=U\=J\=J\=J}  &  12.0      & EWD       & 3        & LWD$_1$   &                  &              &                     &  &  \\
2-29 &  {\tt \=U\=J\=J\=J}  &  12.0      & EWD       & 3        & LWD$_1$   &                  &              &                     &  &  \\
2-30 &  {\tt \=U\=J\=J\=J}  &  12.0      & EWD       & 3        & LWD$_1$   &                  &              &                     &  &  \\
2-31 &  {\tt \=U\=J\=J\=J}  &  12.0      & EWD       & 3        & LWD$_1$   &                  &              &                     &  &  \\
 \hline
\end{tabular}
\end{minipage}
\end{table*}

\begin{table*}
 \centering
 \begin{minipage}{180mm}
 \centering
  \caption{Summary of results for {\tt JJJU} and {\tt \=J\=J\=J\=U}.  See Sec. \ref{tabcol} for a full description of the columns or Table 
          \ref{JUUU} for a summary. \newline
          MS = main sequence, GB = giant branch, EWD = white dwarf with 0-100 Myr cooling, LWD = white dwarf beyond 100 Myr cooling.\newline 
          Notes:\newline $^{\rm a}$Subsequent evolution may have been affected by tides on the GB phase.\newline 
                  $^{\rm b}$Subsequent evolution may have been affected by tides on the MS phase.\newline
                  $^{\rm c}$Subsequent evolution may have been affected by tides on the WD phase.    }
          \label{JJJU}
  \begin{tabular}{@{}ccccccccccc@{}}
 \hline
Sim \#  &  Setup            & $\beta$ & Unpack    & \# Surv & Engulf & Eject        & Collision     & $< R_{\rm max}$? & TPs Eng & Notes   \\
 \hline
3-1  &  {\tt JJJU}          &  6.0    & EWD           & 2             &          & EWD$_3$,EWD$_4$     &                &                 &         &  \\
3-2  &  {\tt JJJU}          &  6.0    & EWD           & 2             &          & EWD$_1$,EWD$_4$     &                &   \#3$_{1.543}$  &         &  \\
3-3  &  {\tt JJJU}          &  6.0    & EWD           & 2             &          & EWD$_2$,EWD$_4$     &                &   \#3$_{0.743}$  &         & a \\
3-4  &  {\tt JJJU}          &  6.0    & EWD           & 2             &          & EWD$_1$,EWD$_4$     &                &                 &         & a \\
 \hline
3-5  &  {\tt JJJU}          &  6.5    & MS          & 2             & MS$_1$   & MS$_4$          &                &                 &         & b \\
3-6  &  {\tt JJJU}          &  6.5    & MS          & 2             &          & MS$_4$,LWD$_3$   &                &   \#1$_{0.505}$  &         &   \\
3-7  &  {\tt JJJU}          &  6.5    & MS          & 2             & MS$_2$   & MS$_4$          &                &                 &         &   \\
3-8  &  {\tt JJJU}          &  6.5    & MS          & 2             & MS$_1$   & MS$_4$          &                &                 &         &   \\
 \hline
3-9  &  {\tt JJJU}          &  7.0    & LWD          & 3             &          & LWD$_4$          &                &                 &         &   \\
3-10 &  {\tt JJJU}          &  7.0    & EWD           & 3             &          & EWD$_4$           &                &                 &         &   \\
3-11 &  {\tt JJJU}          &  7.0    & LWD          & 3             &          & LWD$_4$          &                &                 &         &   \\
3-12 &  {\tt JJJU}          &  7.0    & LWD          & 3             &          & LWD$_4$          &                &                 &         &   \\
 \hline
3-13 &  {\tt JJJU}          &  7.5    &             & 4             &          &                 &                &                 &         &   \\
3-14 &  {\tt JJJU}          &  7.5    &             & 4             &          &                 &                &                 &         &   \\
3-15 &  {\tt JJJU}          &  7.5    &             & 4             &          &                 &                &                 &         &   \\
3-16 &  {\tt JJJU}          &  7.5    &             & 4             &          &                 &                &                 &         &   \\
 \hline
3-17 &  {\tt JJJU}          &  8.0    &             & 4             &          &                 &                &                 &         &   \\
3-18 &  {\tt JJJU}          &  8.0    &             & 4             &          &                 &                &                 &         &   \\
3-19 &  {\tt JJJU}          &  8.0    &             & 4             &          &                 &                &                 &         &   \\
3-20 &  {\tt JJJU}          &  8.0    &             & 4             &          &                 &                &                 &         &   \\
 \hline
3-21 &  {\tt \=J\=J\=J\=U}  &  10.0   & EWD          & 3             & LWD$_4$   &                 &                &                 &         &   \\
3-22 &  {\tt \=J\=J\=J\=U}  &  10.0   & EWD           & 3             & LWD$_4$   &                 &                &                 &         &   \\
3-23 &  {\tt \=J\=J\=J\=U}  &  10.0   & LWD          & 3             & LWD$_4$   &                 &                &                 &         &   \\
3-24 &  {\tt \=J\=J\=J\=U}  &  10.0   & EWD           & 3             & LWD$_4$   &                 &                &                 &         &   \\
 \hline
3-25 &  {\tt \=J\=J\=J\=U}  &  11.0   & LWD          & 4             &          &                 &                & \#2$_{0.515}$, \#4$_{0.00838}$ &   & c  \\
3-26 &  {\tt \=J\=J\=J\=U}  &  11.0   & LWD          & 3             &          &                 &    LWD$_{1-3}$   &                 &         &   \\
3-27 &  {\tt \=J\=J\=J\=U}  &  11.0   & LWD          & 3             & LWD$_4$   &                 &                &                 &         &   \\
3-28 &  {\tt \=J\=J\=J\=U}  &  11.0   & LWD          & 3             & LWD$_4$   &                 &                &                 &         &   \\
 \hline
3-29 &  {\tt \=J\=J\=J\=U}  &  12.0   & EWD           & 3             & LWD$_4$   &                 &                &                 &         &   \\
3-30 &  {\tt \=J\=J\=J\=U}  &  12.0   & EWD           & 3             & LWD$_4$   &                 &                &                 &         &   \\
3-31 &  {\tt \=J\=J\=J\=U}  &  12.0   & MS          & 2             &          &  LWD$_4$         &    MS$_{1-2}$   &                 &         &   \\
3-32 &  {\tt \=J\=J\=J\=U}  &  12.0   & MS          & 3             & LWD$_4$   &                 &                &                 &         & b  \\
\hline
\end{tabular}
\end{minipage}
\end{table*}

\begin{table*}
 \centering
 \begin{minipage}{180mm}
 \centering
  \caption{Summary of results for {\tt UUUJ} and {\tt \=U\=U\=U\=J}.  See Sec. \ref{tabcol} for a full description of the columns or Table 
          \ref{JUUU} for a summary. \newline
          MS = main sequence, GB = giant branch, EWD = white dwarf with 0-100 Myr cooling, LWD = white dwarf beyond 100 Myr cooling.\newline
          Notes:\newline 
                  $^{\rm d}$A test particle which survived the entire integration acheived a minimum pericentre of 0.0244 au at a WD cooling age of 1.844 Gyr.\newline
                  $^{\rm e}$A test particle which survived the entire integration acheived a minimum pericentre of 0.5537 au at a WD cooling age of 4.446 Gyr.\newline
                  $^{\rm f}$A test particle which survived the entire integration acheived a minimum pericentre of 0.2312 au at a WD cooling age of 6.671 Gyr.\newline
    }
          \label{UUUJ}
  \begin{tabular}{@{}ccccccccccc@{}}
 \hline
Sim \#  &  Setup            & $\beta$ & Unpack    & \# Surv & Engulf & Eject       & Collision     & $< R_{\rm max}$? & TPs Eng & Notes   \\
 \hline
4-1  &  {\tt UUUJ}          &  7.0    & MS          & 2             &          & MS$_1$,MS$_2$   &                &                 &         &  \\
4-2  &  {\tt UUUJ}          &  7.0    & MS          & 2             &          & MS$_1$,MS$_2$   &                &                 &         &  \\
4-3  &  {\tt UUUJ}          &  7.0    & MS          & 2             &          & MS$_3$          &  MS$_{1-2}$     &                 &         &  \\
4-4  &  {\tt UUUJ}          &  7.0    & MS          & 1             &   GB$_1$ & MS$_2$,MS$_3$   &                &                 &         &  \\
\hline
4-5  &  {\tt UUUJ}          &  7.5    & MS          & 2             &          & MS$_2$,MS$_3$   &                &                 &         &  \\
4-6  &  {\tt UUUJ}          &  7.5    & MS          & 2             &          & MS$_2$,MS$_3$   &                &                 &         &  \\
4-7  &  {\tt UUUJ}          &  7.5    & MS          & 2             &          & MS$_3$          &  MS$_{1-4}$     &                 &         &  \\
4-8  &  {\tt UUUJ}          &  7.5    & MS          & 2             &          & GB$_3$          &  MS$_{1-2}$     &                 &         &  \\
\hline
4-9  &  {\tt UUUJ}          &  8.0    & MS          & 2             &          & MS$_1$,WD$_2$   &                &                 &         &  \\
4-10 &  {\tt UUUJ}          &  8.0    & MS          & 2             &          & MS$_1$,MS$_2$   &                &                 &         &  \\
4-11 &  {\tt UUUJ}          &  8.0    & MS          & 2             &          & MS$_2$,MS$_3$   &                &                 &         &  \\
4-12 &  {\tt UUUJ}          &  8.0    & MS          & 2             &          & MS$_1$,MS$_3$   &                &                 &         &  \\
\hline
4-13 &  {\tt UUUJ}          &  8.5    & EWD           & 2             &          & EWD$_2$,EWD$_3$     &                &                 &  (8/0)  &  \\
4-14 &  {\tt UUUJ}          &  8.5    & EWD           & 2             &          & EWD$_2$,EWD$_3$     &                &                 &  (12/0) &  \\
4-15 &  {\tt UUUJ}          &  8.5    & EWD           & 2             &          & EWD$_2$,EWD$_3$     &                &                 &  (12/0) &  \\
4-16 &  {\tt UUUJ}          &  8.5    & EWD          & 2             &          & EWD$_2$           & EWD$_{1-3}$       &                 &  (3/1)  &  \\
\hline
4-17 &  {\tt UUUJ}          &  9.0    & LWD          & 2             &          & LWD$_2$,LWD$_3$   &                &                 &  (0/12) &  \\
4-18 &  {\tt UUUJ}          &  9.0    & LWD          & 2             &          & LWD$_2$,LWD$_3$   &                &   \#1$_{0.464}$  &  (0/12) &  \\
4-19 &  {\tt UUUJ}          &  9.0    & LWD          & 2             &          & LWD$_1$,LWD$_3$   &                &                 &  (0/10) &  \\
4-20 &  {\tt UUUJ}          &  9.0    & LWD          & 2             &          & LWD$_2$,LWD$_3$   &                &                 &  (0/7)  &  \\
\hline
4-21 &  {\tt UUUJ}          &  9.5    & LWD          & 2             &          & LWD$_1$,LWD$_2$   &                &                 &  (0/11) &  \\
4-22 &  {\tt UUUJ}          &  9.5    & LWD          & 2             &          & LWD$_1$,LWD$_2$   &                &                 &  (0/11) & d \\
4-23 &  {\tt UUUJ}          &  9.5    & LWD          & 2             &          & LWD$_1$,LWD$_3$   &                &                 &  (0/11) &  \\
4-24 &  {\tt UUUJ}          &  9.5    & LWD          & 2             &          & LWD$_1$,LWD$_3$   &                &                 &  (0/10) & e,f \\
\hline
4-25 &  {\tt \=U\=U\=U\=J}  &  10.0   & MS          & 2             &          &                 & MS$_{2-4}$,LWD$_{3-4}$    &         &         &   \\
4-26 &  {\tt \=U\=U\=U\=J}  &  10.0   & MS          & 3             &          &                 & MS$_{2-4}$,    &                 &         &   \\
4-27 &  {\tt \=U\=U\=U\=J}  &  10.0   & MS          & 2             &          &                 & MS$_{1-4}$,MS$_{2-3}$    &         &         &   \\
4-28 &  {\tt \=U\=U\=U\=J}  &  10.0   & MS          & 2             &          &                 & MS$_{1-4}$,MS$_{2-4}$    &         &         &   \\
\hline
4-29 &  {\tt \=U\=U\=U\=J}  &  12.0   & MS          & 4             &          &                 &               &                 &         &   \\
4-30 &  {\tt \=U\=U\=U\=J}  &  12.0   & MS          & 3             &          &                 & MS$_{1-4}$     &                 &         &   \\
4-31 &  {\tt \=U\=U\=U\=J}  &  12.0   & MS          & 2             &          &                 & MS$_{2-4}$,MS$_{3-4}$    &         &         &   \\
4-32 &  {\tt \=U\=U\=U\=J}  &  12.0   & MS          & 3             &          &                 & MS$_{2-4}$     &         &         &   \\
\hline
4-33 &  {\tt \=U\=U\=U\=J}  &  14.0   & LWD          & 4             &          &                 &               &                 &         &   \\
4-34 &  {\tt \=U\=U\=U\=J}  &  14.0   & LWD          & 4             &          &                 &               &                 &         &   \\
4-35 &  {\tt \=U\=U\=U\=J}  &  14.0   & LWD          & 4             &          &                 &               &                 &         &   \\
4-36 &  {\tt \=U\=U\=U\=J}  &  14.0   & LWD          & 4             &          &                 &               &                 &         &   \\
\hline
\end{tabular}
\end{minipage}
\end{table*}

\begin{table*}
 \centering
 \begin{minipage}{180mm}
 \centering
  \caption{Summary of results for {\tt JUUJ} and {\tt \=J\=U\=U\=J}.  See Sec. \ref{tabcol} for a full description of the columns or Table 
          \ref{JUUU} for a summary. \newline
          MS = main sequence, GB = giant branch, EWD = white dwarf with 0-100 Myr cooling, LWD = white dwarf beyond 100 Myr cooling.\newline
          Notes:\newline 
                  $^{\rm b}$Subsequent evolution may have been affected by tides on the MS phase.\newline
                  $^{\rm c}$Subsequent evolution may have been affected by tides on the WD phase.\newline   
                  $^{\rm z}$Unpacking, ejections and engulfments all occur for WD cooling ages exceeding 10 Gyr.}
          \label{JUUJ}
  \begin{tabular}{@{}ccccccccccc@{}}
 \hline
Sim \#  &  Setup            & $\beta$ & Unpack    & \# Surv & Engulf & Eject        & Collision     & $< R_{\rm max}$? &  TPs Eng & Notes   \\
 \hline
5-1  &  {\tt JUUJ}          &  6.0    & MS          & 2             &          & MS$_2$,MS$_3$   &                &                 &         &  \\
5-2  &  {\tt JUUJ}          &  6.0    & MS          & 2             &          & MS$_2$,MS$_3$   &                &                 &         &  \\
5-3  &  {\tt JUUJ}          &  6.0    & MS          & 2             &          & MS$_2$,MS$_3$   &                &                 &         &  \\
5-4  &  {\tt JUUJ}          &  6.0    & MS          & 2             &          & MS$_2$,MS$_3$   &                &                 &         &  \\
\hline
5-5  &  {\tt JUUJ}          &  6.5    & MS          & 2             &          & MS$_3$,EWD$_2$   &                &                 &         &  \\
5-6  &  {\tt JUUJ}          &  6.5    & MS          & 2             &          & MS$_2$,MS$_3$  &                &                 &         &  \\
5-7  &  {\tt JUUJ}          &  6.5    & MS          & 2             &          & MS$_2$,MS$_3$  &                &                 &         &  \\
5-8  &  {\tt JUUJ}          &  6.5    & MS          & 2             &          & MS$_2$,MS$_3$  &                &                 &         &  \\
\hline
5-9  &  {\tt JUUJ}          &  7.0    & LWD          & 2             &          & LWD$_2$,LWD$_3$  &                &                 &         &  \\
5-10 &  {\tt JUUJ}          &  7.0    & MS          & 2             &          & MS$_2$,MS$_3$  &                &                 &         &  \\
5-11 &  {\tt JUUJ}          &  7.0    & EWD           & 2             &          & EWD$_2$,EWD$_3$    &                &                 &         &  \\
5-12 &  {\tt JUUJ}          &  7.0    & MS          & 2             &          & MS$_2$,EWD$_3$   &                &                 &         &  \\
\hline
5-13 &  {\tt JUUJ}          &  7.5    & MS          & 2             &          & MS$_2$,MS$_3$  &                &                 &         &  \\
5-14 &  {\tt JUUJ}          &  7.5    & MS          & 2             &          & MS$_2$         &    MS$_{1-3}$   &                 &         &  \\
5-15 &  {\tt JUUJ}          &  7.5    & MS          & 2             &          & MS$_2$,MS$_3$  &                &                 &         &  \\
5-16 &  {\tt JUUJ}          &  7.5    & MS          & 2             &          & MS$_2$,MS$_3$  &                &                 &         &  \\
\hline
5-17 &  {\tt JUUJ}          &  8.0    & MS          & 2             &          & MS$_2$,MS$_3$  &                &                 &         &  \\
5-18 &  {\tt JUUJ}          &  8.0    & MS          & 2             &   MS$_2$ & MS$_3$         &                &                 &         &  \\
5-19 &  {\tt JUUJ}          &  8.0    & MS          & 2             &          & MS$_2$,MS$_3$  &                &                 &         &  \\
5-20 &  {\tt JUUJ}          &  8.0    & MS          & 2             &          & MS$_2$,MS$_3$  &                &                 &         &  \\
\hline
5-21 &  {\tt JUUJ}          &  8.5    & EWD           & 2             &          & EWD$_2$,EWD$_3$    &                &                 &   (2/0)   &  \\
5-22 &  {\tt JUUJ}          &  8.5    & MS          & 2             &          & MS$_2$,MS$_3$  &                &                 &   (0/0)   &  \\
5-23 &  {\tt JUUJ}          &  8.5    & EWD           & 2             &          & EWD$_2$,EWD$_3$    &                &                 &   (0/0)   &  \\
5-24 &  {\tt JUUJ}          &  8.5    & EWD           & 2             &          & EWD$_2$,EWD$_3$    &                &                 &   (0/0)   &  \\
\hline
5-25 &  {\tt JUUJ}          &  9.0    & EWD           & 2             &          & EWD$_2$,EWD$_3$    &                &                 &   (0/0)   &  \\
5-26 &  {\tt JUUJ}          &  9.0    & EWD           & 2             &          & EWD$_2$,EWD$_3$    &                &                 &   (0/0)   &  \\
5-27 &  {\tt JUUJ}          &  9.0    & EWD           & 2             &          & EWD$_2$,LWD$_3$   &                &                 &   (0/2)   &  \\
5-28 &  {\tt JUUJ}          &  9.0    & EWD           & 2             &          & EWD$_2$,EWD$_3$    &                &                 &   (4/0)   &  \\
\hline
5-29 &  {\tt JUUJ}          &  9.5    & LWD          & 2             &          & LWD$_2$,LWD$_3$  &                &                 &   (0/3)   &  \\
5-30 &  {\tt JUUJ}          &  9.5    &             & 4             &          &                &                &                 &   (0/0)   &  \\
5-31 &  {\tt JUUJ}          &  9.5    & LWD          & 2             &          & LWD$_2$,LWD$_3$  &                &                 &   (0/2)   &  \\
5-32 &  {\tt JUUJ}          &  9.5    & LWD          & 2             &          & LWD$_2$,LWD$_3$  &                &                 &   (0/3)   & z \\
\hline
5-33 &  {\tt \=J\=U\=U\=J}  &  10.0   & MS          & 3             &          &                &    MS$_{1-2}$   &                 &          &  \\
5-34 &  {\tt \=J\=U\=U\=J}  &  10.0   & MS          & 3             &          &                &    LWD$_{1-4}$   &                 &          &  \\
5-35 &  {\tt \=J\=U\=U\=J}  &  10.0   & MS          & 2             &          & LWD$_2$         &    LWD$_{1-4}$   &                 &          &  \\
5-36 &  {\tt \=J\=U\=U\=J}  &  10.0   & MS          & 4             &          &                &                &    \#3$_{1.624}$ &          &  \\
\hline
5-37 &  {\tt \=J\=U\=U\=J}  &  12.0   & LWD          & 3             &          & LWD$_2$         &                &                 &          &  \\
5-38 &  {\tt \=J\=U\=U\=J}  &  12.0   & MS          & 3             &          &                &   GB$_{2-4}$    &                 &          &  \\
5-39 &  {\tt \=J\=U\=U\=J}  &  12.0   & MS          & 3             &          &                &   GB$_{1-2}$    &                 &          &  \\
5-40 &  {\tt \=J\=U\=U\=J}  &  12.0   & EWD           & 4             &          &                &                &                 &          &  \\
\hline
5-41 &  {\tt \=J\=U\=U\=J}  &  14.0   & LWD          & 3             &          &                &    LWD$_{1-2}$   &                 &          &  \\
5-42 &  {\tt \=J\=U\=U\=J}  &  14.0   & LWD          & 4             &          &                &                &                 &          &  \\
5-43 &  {\tt \=J\=U\=U\=J}  &  14.0   & LWD          & 4             &          &                &                &                 &          &  \\
5-44 &  {\tt \=J\=U\=U\=J}  &  14.0   &             & 4             &          &                &                &                 &          &  \\
\hline
\end{tabular}
\end{minipage}
\end{table*}

\begin{table*}
 \centering
 \begin{minipage}{180mm}
 \centering
  \caption{Summary of results for {\tt UJJU} and {\tt \=U\=J\=J\=U}.  See Sec. \ref{tabcol} for a full description of the columns or Table 
          \ref{JUUU} for a summary. \newline
          MS = main sequence, GB = giant branch, EWD = white dwarf with 0-100 Myr cooling, LWD = white dwarf beyond 100 Myr cooling.\newline
          }
          \label{UJJU}
  \begin{tabular}{@{}ccccccccccc@{}}
 \hline
Sim \#  &  Setup            & $\beta$ & Unpack    & \# Surv & Engulf & Eject        & Collision     & $< R_{\rm max}$? &  TPs Eng & Notes   \\
 \hline
6-1  &  {\tt UJJU}          &  6.0    & EWD           & 2             &          & EWD$_1$,EWD$_4$   &                &                 &         &  \\
6-2  &  {\tt UJJU}          &  6.0    & EWD           & 2             &          & EWD$_1$,EWD$_4$   &                &                 &         &  \\
6-3  &  {\tt UJJU}          &  6.0    & EWD           & 2             & EWD$_1$    & EWD$_4$         &                &                 &         &  \\
6-4  &  {\tt UJJU}          &  6.0    & EWD           & 2             &          & EWD$_1$,EWD$_4$   &                &                 &         &  \\
\hline
6-5  &  {\tt UJJU}          &  6.5    & MS          & 2             &          & MS$_1$,MS$_4$ &                &                 &         &  \\
6-6  &  {\tt UJJU}          &  6.5    & MS          & 2             &          & MS$_1$,MS$_4$ &                &                 &         &  \\
6-7  &  {\tt UJJU}          &  6.5    & MS          & 2             & MS$_1$   & MS$_4$        &                &                 &         &  \\
6-8  &  {\tt UJJU}          &  6.5    & MS          & 2             &          & MS$_1$,MS$_4$ &                &                 &         &  \\
\hline
6-9  &  {\tt UJJU}          &  7.0    &             & 4             &          &               &                &                 &         &  \\
6-10 &  {\tt UJJU}          &  7.0    & LWD          & 2             &          & LWD$_1$,LWD$_4$ &                &                 &         &  \\
6-11 &  {\tt UJJU}          &  7.0    &             & 4             &          &               &                &                 &         &  \\
6-12 &  {\tt UJJU}          &  7.0    &             & 4             &          &               &                &                 &         &  \\
\hline
6-13 &  {\tt UJJU}          &  7.5    &             & 4             &          &               &                &                 &         &  \\
6-14 &  {\tt UJJU}          &  7.5    &             & 4             &          &               &                &                 &         &  \\
6-15 &  {\tt UJJU}          &  7.5    &             & 4             &          &               &                &                 &         &  \\
6-16 &  {\tt UJJU}          &  7.5    &             & 4             &          &               &                &                 &         &  \\
\hline
6-17 &  {\tt UJJU}          &  8.0    &             & 4             &          &               &                &                 &         &  \\
6-18 &  {\tt UJJU}          &  8.0    &             & 4             &          &               &                &                 &         &  \\
6-19 &  {\tt UJJU}          &  8.0    &             & 4             &          &               &                &                 &         &  \\
6-20 &  {\tt UJJU}          &  8.0    &             & 4             &          &               &                &                 &         &  \\
\hline
6-21 &  {\tt \=U\=J\=J\=U}  &  9.0    & EWD           & 3             &          &               &  LWD$_{2-3}$     &                 &         &  \\
6-22 &  {\tt \=U\=J\=J\=U}  &  9.0    & MS          & 3             &          &               &  MS$_{1-3}$     &                 &         &  \\
6-23 &  {\tt \=U\=J\=J\=U}  &  9.0    & LWD          & 3             &          &               &  LWD$_{2-3}$     &                 &         &  \\
6-24 &  {\tt \=U\=J\=J\=U}  &  9.0    & MS          & 2             &          & LWD$_1$        &  MS$_{2-4}$     &                 &         &  \\
\hline
6-25 &  {\tt \=U\=J\=J\=U}  & 10.0    & LWD          & 4             &          &               &                &                 &         &  \\
6-26 &  {\tt \=U\=J\=J\=U}  & 10.0    & LWD          & 4             &          &               &                &                 &         &  \\
6-27 &  {\tt \=U\=J\=J\=U}  & 10.0    & LWD          & 4             &          &               &                &                 &         &  \\
6-28 &  {\tt \=U\=J\=J\=U}  & 10.0    &             & 4             &          &               &                &                 &         &  \\
\hline
6-29 &  {\tt \=U\=J\=J\=U}  & 11.0    & LWD          & 4             &          &               &                &                 &         &  \\
6-30 &  {\tt \=U\=J\=J\=U}  & 11.0    & LWD          & 3             &          & LWD$_4$        &                &                 &         &  \\
6-31 &  {\tt \=U\=J\=J\=U}  & 11.0    & EWD           & 4             &          &               &                &                 &         &  \\
6-32 &  {\tt \=U\=J\=J\=U}  & 11.0    & LWD          & 4             &          &               &                &  \#1$_{1.696}$   &         &  \\
\hline
\end{tabular}
\end{minipage}
\end{table*}

\begin{table*}
 \centering
 \begin{minipage}{180mm}
 \centering
  \caption{Summary of results for {\tt JSUN} and {\tt \=J\=S\=U\=N}.  See Sec. \ref{tabcol} for a full description of the columns or Table 
          \ref{JUUU} for a summary. \newline
          MS = main sequence, GB = giant branch, EWD = white dwarf with 0-100 Myr cooling, LWD = white dwarf beyond 100 Myr cooling.\newline 
          Notes:\newline $^{\rm a}$Subsequent evolution may have been affected by tides on the GB phase.\newline 
          }
          \label{JSUN}
  \begin{tabular}{@{}ccccccccccc@{}}
 \hline
Sim \#  &  Setup            & $\beta$ & Unpack    & \# Surv & Engulf & Eject        & Collision     & $< R_{\rm max}$? &  TPs Eng & Notes   \\
 \hline
7-1  &  {\tt JSUN}          &  6.0    & MS          & 2             &          & MS$_3$,MS$_4$   &                &                 &         &  \\
7-2  &  {\tt JSUN}          &  6.0    & MS          & 2             &          & MS$_3$,WD$_4$   &                &                 &         &  \\
7-3  &  {\tt JSUN}          &  6.0    & MS          & 2             &          & MS$_3$,MS$_4$   &                &                 &         &  \\
7-4  &  {\tt JSUN}          &  6.0    & EWD           & 2             &          & EWD$_3$,EWD$_4$     &                &                 &         &  \\
\hline
7-5  &  {\tt JSUN}          &  6.5    & MS          & 2             &          & MS$_3$,MS$_4$   &                &                 &         &  \\
7-6  &  {\tt JSUN}          &  6.5    & EWD           & 1             &          & EWD$_3$,EWD$_4$,LWD$_2$   &           &                 &         &  \\
7-7  &  {\tt JSUN}          &  6.5    & MS          & 2             &          & MS$_3$,MS$_4$   &                &                 &         &  \\
7-8  &  {\tt JSUN}          &  6.5    & MS          & 2             &          & MS$_3$,MS$_4$   &                &                 &         &  \\
\hline
7-9  &  {\tt JSUN}          &  7.0    & MS          & 2             & MS$_4$   & MS$_3$          &                &                 &         &  \\
7-10 &  {\tt JSUN}          &  7.0    & MS          & 2             &          & EWD$_4$           &  MS$_{3-4}$    &                 &         &  \\
7-11 &  {\tt JSUN}          &  7.0    & MS          & 1             &          & MS$_3$,MS$_4$,LWD$_2$   &         &                 &         &  \\
7-12 &  {\tt JSUN}          &  7.0    & MS          & 2             &          & MS$_3$,MS$_4$   &               &                 &         &  \\
\hline
7-13 &  {\tt JSUN}          &  7.5    & MS          & 2             &          & MS$_3$,MS$_4$   &               &                 &         &  \\
7-14 &  {\tt JSUN}          &  7.5    & MS          & 2             &          & MS$_4$          &   MS$_{1-3}$   &                 &         &  \\
7-15 &  {\tt JSUN}          &  7.5    & MS          & 2             &          & MS$_3$,MS$_4$   &               &                 &         &  \\
7-16 &  {\tt JSUN}          &  7.5    & MS          & 2             &          & MS$_3$,LWD$_4$   &               &                 &         &  \\
\hline
7-17 &  {\tt JSUN}          &  8.0    & EWD           & 2             &          & EWD$_3$,LWD$_4$    &               &                 &         &  \\
7-18 &  {\tt JSUN}          &  8.0    & EWD          & 2             &          & EWD$_3$,EWD$_4$     &               &                 &         &  \\
7-19 &  {\tt JSUN}          &  8.0    & LWD          & 2             &          & LWD$_3$,LWD$_4$   &               &                 &         &  \\
7-20 &  {\tt JSUN}          &  8.0    & EWD           & 1             &          & EWD$_3$,LWD$_2$,LWD$_4$   &         &                 &         &  \\
\hline
7-21 &  {\tt JSUN}          &  8.5    &             & 4             &          &                 &              &                 &   (0/0)  &  \\
7-22 &  {\tt JSUN}          &  8.5    & LWD          & 2             &          & LWD$_3$,LWD$_4$   &              &                 &   (0/2)  &  \\
7-23 &  {\tt JSUN}          &  8.5    & LWD          & 3             &          & LWD$_3$          &              &                 &   (0/3)  &  \\
7-24 &  {\tt JSUN}          &  8.5    &             & 4             &          &                 &              &                 &   (0/0)  &  \\
\hline
7-25 &  {\tt JSUN}          &  9.0    &             & 4             &          &                 &              &                 &   (0/0)  &  \\
7-26 &  {\tt JSUN}          &  9.0    &             & 4             &          &                 &              &                 &   (0/0)  &  \\
7-27 &  {\tt JSUN}          &  9.0    &             & 4             &          &                 &              &                 &   (0/0)  &  \\
7-28 &  {\tt JSUN}          &  9.0    &             & 4             &          &                 &              &                 &   (0/0)  &  \\
\hline
7-29 &  {\tt JSUN}          &  9.5    &             & 4             &          &                 &              &                 &   (0/0)  &  \\
7-30 &  {\tt JSUN}          &  9.5    &             & 4             &          &                 &              &                 &   (0/0)  &  \\
7-31 &  {\tt JSUN}          &  9.5    &             & 4             &          &                 &              &                 &   (0/0)  &  \\
7-32 &  {\tt JSUN}          &  9.5    &             & 4             &          &                 &              &                 &   (0/0)  &  \\
\hline
7-33 &  {\tt \=J\=S\=U\=N}  & 10.0    & MS          & 2             &          & MS$_3$,LWD$_2$    &              &                 &          &  \\
7-34 &  {\tt \=J\=S\=U\=N}  & 10.0    & MS          & 4             &          &                 &              &                 &          &  \\
7-35 &  {\tt \=J\=S\=U\=N}  & 10.0    & MS          & 3             &          &                 &   MS$_{2-4}$  &                 &          &  \\
7-36 &  {\tt \=J\=S\=U\=N}  & 10.0    & MS          & 2             &          &                 & MS$_{1-2}$,LWD$_{1-4}$  &         &          &  \\
\hline
7-37 &  {\tt \=J\=S\=U\=N}  & 11.0    & MS          & 4             &          &                 &              &                 &          &  \\
7-38 &  {\tt \=J\=S\=U\=N}  & 11.0    & MS          & 3             &          &                 &   MS$_{2-4}$  &                 &          &  \\
7-39 &  {\tt \=J\=S\=U\=N}  & 11.0    & MS          & 4             &          &                 &   MS$_{2-4}$  &                 &          &  \\
7-40 &  {\tt \=J\=S\=U\=N}  & 11.0    & EWD           & 4             &          &                 &              &                 &          &  \\
\hline
7-41 &  {\tt \=J\=S\=U\=N}  & 12.0    & MS          & 3             &          &                 &   LWD$_{1-4}$  &                 &          &  \\
7-42 &  {\tt \=J\=S\=U\=N}  & 12.0    & MS          & 3             &          &                 &   MS$_{1-3}$  &                 &          &  \\
7-43 &  {\tt \=J\=S\=U\=N}  & 12.0    & MS          & 3             &          &                 &   MS$_{1-4}$  &                 &          &  \\
7-44 &  {\tt \=J\=S\=U\=N}  & 12.0    & MS          & 4             &          &                 &              & \#3$_{0.914}$, \#4$_{0.455}$  & & a \\
\hline
\end{tabular}
\end{minipage}
\end{table*}

\begin{table*}
 \centering
 \begin{minipage}{180mm}
 \centering
  \caption{Summary of results for {\tt JSNU} and {\tt \=J\=S\=N\=U}.  See Sec. \ref{tabcol} for a full description of the columns or Table 
          \ref{JUUU} for a summary. \newline
          MS = main sequence, GB = giant branch, EWD = white dwarf with 0-100 Myr cooling, LWD = white dwarf beyond 100 Myr cooling.\newline
          }
          \label{JSNU}
  \begin{tabular}{@{}ccccccccccc@{}}
 \hline
Sim \#  &  Setup            & $\beta$ & Unpack    & \# Surv & Engulf & Eject        & Collision     & $< R_{\rm max}$? &  TPs Eng & Notes   \\
 \hline
8-1  &  {\tt JSNU}          &  6.0    & MS          & 2             &          & MS$_3$,MS$_4$   &                &                 &         &  \\
8-2  &  {\tt JSNU}          &  6.0    & MS          & 2             &          & MS$_3$,MS$_4$   &                &                 &         &  \\
8-3  &  {\tt JSNU}          &  6.0    & MS          & 2             &          & MS$_3$,MS$_4$   &                &                 &         &  \\
8-4  &  {\tt JSNU}          &  6.0    & MS          & 2             &          & MS$_3$,MS$_4$   &                &                 &         &  \\
\hline
8-5  &  {\tt JSNU}          &  6.5    & MS          & 2             &          & MS$_2$,MS$_4$   &                &                 &         &  \\
8-6  &  {\tt JSNU}          &  6.5    & MS          & 2             &          & MS$_3$,MS$_4$   &                &                 &         &  \\
8-7  &  {\tt JSNU}          &  6.5    & EWD           & 2             &          & EWD$_3$,EWD$_4$     &                &                 &         &  \\
8-8  &  {\tt JSNU}          &  6.5    & EWD           & 2             &          & EWD$_3$,EWD$_4$     &                &                 &         &  \\
\hline
8-9  &  {\tt JSNU}          &  7.0    & MS          & 2             &          & MS$_3$,MS$_4$   &                &                 &         &  \\
8-10 &  {\tt JSNU}          &  7.0    & MS          & 2             &          & MS$_3$,MS$_4$   &                &                 &         &  \\
8-11 &  {\tt JSNU}          &  7.0    & MS          & 2             &          & MS$_3$,EWD$_4$    &                &                 &         &  \\
8-12 &  {\tt JSNU}          &  7.0    & MS          & 2             &          & MS$_3$,MS$_4$   &                &                 &         &  \\
\hline
8-13 &  {\tt JSNU}          &  7.5    & MS          & 2             &   MS$_4$ & MS$_3$          &                &                 &         &  \\
8-14 &  {\tt JSNU}          &  7.5    & MS          & 2             &          & MS$_3$,MS$_4$   &                &                 &         &  \\
8-15 &  {\tt JSNU}          &  7.5    & MS          & 2             &          & MS$_3$          &    MS$_{1-4}$   &                 &         &  \\
8-16 &  {\tt JSNU}          &  7.5    & MS          & 2             &          & MS$_3$,MS$_4$   &                &                 &         &  \\
\hline
8-17 &  {\tt JSNU}          &  8.0    & LWD          & 2             &          & LWD$_2$,LWD$_4$   &                &                 &         &  \\
8-18 &  {\tt JSNU}          &  8.0    & LWD          & 2             &          & LWD$_3$,LWD$_4$   &                &                 &         &  \\
8-19 &  {\tt JSNU}          &  8.0    & EWD           & 2             &          & LWD$_3$,LWD$_4$   &                &                 &         &  \\
8-20 &  {\tt JSNU}          &  8.0    & EWD           & 2             &          & EWD$_4$,LWD$_3$    &                &                 &         &  \\
\hline
8-21 &  {\tt JSNU}          &  8.5    & LWD          & 2             &          & LWD$_3$,LWD$_4$   &                &                 &  (0/0)  &  \\
8-22 &  {\tt JSNU}          &  8.5    & LWD          & 2             &          & LWD$_3$,LWD$_4$   &                &                 &  (0/8)  &  \\
8-23 &  {\tt JSNU}          &  8.5    & LWD          & 2             &          & LWD$_3$,LWD$_4$   &                &                 &  (0/1)  &  \\
8-24 &  {\tt JSNU}          &  8.5    & EWD           & 1             &          & LWD$_2$,LWD$_3$,LWD$_4$   &          &                &  (4/2)  &  \\
\hline
8-25 &  {\tt JSNU}          &  9.0    &             & 4             &          &                &                &                 &  (0/0)  &  \\
8-26 &  {\tt JSNU}          &  9.0    &             & 4             &          &                &                &                 &  (0/0)  &  \\
8-27 &  {\tt JSNU}          &  9.0    &             & 4             &          &                &                &                 &  (0/0)  &  \\
8-28 &  {\tt JSNU}          &  9.0    &             & 4             &          &                &                &                 &  (0/0)  &  \\
\hline
8-29 &  {\tt JSNU}          &  9.5    &             & 4             &          &                &                &                 &  (0/0)  &  \\
8-30 &  {\tt JSNU}          &  9.5    &             & 4             &          &                &                &                 &  (0/0)  &  \\
8-31 &  {\tt JSNU}          &  9.5    &             & 4             &          &                &                &                 &  (0/0)  &  \\
8-32 &  {\tt JSNU}          &  9.5    &             & 4             &          &                &                &                 &  (0/0)  &  \\
\hline
8-33 &  {\tt \=J\=S\=N\=U}  &  10.0   & MS          & 3             &          &                &    LWD$_{1-2}$   &                 &         &  \\
8-34 &  {\tt \=J\=S\=N\=U}  &  10.0   & MS          & 3             &          &                &    LWD$_{1-2}$   &                 &         &  \\
8-35 &  {\tt \=J\=S\=N\=U}  &  10.0   & MS          & 2             &          &                &  MS$_{1-2}$,MS$_{1-3}$   &         &         &  \\
8-36 &  {\tt \=J\=S\=N\=U}  &  10.0   & MS          & 3             &          &                &    MS$_{2-4}$   &                 &         &  \\
\hline
8-37 &  {\tt \=J\=S\=N\=U}  &  11.0   & MS          & 2             &          &                &  MS$_{1-2}$,MS$_{1-3}$   &         &         &  \\
8-38 &  {\tt \=J\=S\=N\=U}  &  11.0   & MS          & 3             &          & LWD$_4$         &                &                  &         &  \\
8-39 &  {\tt \=J\=S\=N\=U}  &  11.0   & MS          & 3             &          &                &  MS$_{1-3}$     &                  &         &  \\
8-40 &  {\tt \=J\=S\=N\=U}  &  11.0   & MS          & 3             &          &                &  MS$_{1-2}$     &                  &         &  \\
\hline
8-41 &  {\tt \=J\=S\=N\=U}  &  12.0   & MS          & 4             &          &                &                &                  &         &  \\
8-42 &  {\tt \=J\=S\=N\=U}  &  12.0   & MS          & 3             &          &                &  GB$_{1-2}$     &                  &         &  \\
8-43 &  {\tt \=J\=S\=N\=U}  &  12.0   & EWD           & 4             &          &                &                &     \#3$_{0.835}$ &         &  \\
8-44 &  {\tt \=J\=S\=N\=U}  &  12.0   & MS          & 3             &          &                &   MS$_{1-2}$    &                  &         &  \\
\hline
\end{tabular}
\end{minipage}
\end{table*}

\begin{table*}
 \centering
 \begin{minipage}{180mm}
 \centering
  \caption{Summary of results for {\tt UNJS} and {\tt \=U\=N\=J\=S}.  See Sec. \ref{tabcol} for a full description of the columns or Table 
          \ref{JUUU} for a summary. \newline
          MS = main sequence, GB = giant branch, EWD = white dwarf with 0-100 Myr cooling, LWD = white dwarf beyond 100 Myr cooling.\newline
          Notes:\newline $^{\rm a}$Subsequent evolution may have been affected by tides on the GB phase.}
          \label{UNJS}
  \begin{tabular}{@{}ccccccccccc@{}}
 \hline
Sim \#  &  Setup            & $\beta$ & Unpack    & \# Surv & Engulf & Eject        & Collision     & $< R_{\rm max}$? &  TPs Eng & Notes   \\
 \hline
9-1  &  {\tt UNJS}          &  6.0    & MS          & 2             &          & MS$_1$,LWD$_2$   &                &                 &         &  \\
9-2  &  {\tt UNJS}          &  6.0    & MS          & 2             &          & MS$_1$,MS$_2$   &                &                 &         &  \\
9-3  &  {\tt UNJS}          &  6.0    & MS          & 2             &          & MS$_1$,MS$_2$   &                &                 &         &  \\
9-4  &  {\tt UNJS}          &  6.0    & MS          & 3             &          &                 &   MS$_{3-4}$    &                 &         &  \\
\hline
9-5  &  {\tt UNJS}          &  6.5    & EWD           & 2             &          & EWD$_1$,EWD$_2$     &                &                 &         &  \\
9-6  &  {\tt UNJS}          &  6.5    & MS          & 2             &          & MS$_1$,MS$_2$   &                &                 &         &  \\
9-7  &  {\tt UNJS}          &  6.5    & MS          & 2             &          & MS$_1$,MS$_2$   &                &                 &         &  \\
9-8  &  {\tt UNJS}          &  6.5    & EWD           & 2             &          & EWD$_1$,LWD$_2$    &                &                 &         &  \\
\hline
9-9  &  {\tt UNJS}          &  7.0    & EWD           & 2             &          & EWD$_1$,EWD$_2$     &                &                 &         &  \\
9-10 &  {\tt UNJS}          &  7.0    & EWD           & 2             &          & EWD$_1$,EWD$_2$     &                &                 &         &  \\
9-11 &  {\tt UNJS}          &  7.0    & EWD           & 2             &          & EWD$_1$,EWD$_2$     &                &                 &         &  \\
9-12 &  {\tt UNJS}          &  7.0    & EWD           & 2             &          & EWD$_2$,LWD$_1$    &                &                 &         &  \\
\hline
9-13 &  {\tt UNJS}          &  7.5    & MS          & 2             &          & MS$_2$,LWD$_4$   &                &                 &         &  \\
9-14 &  {\tt UNJS}          &  7.5    & MS          & 3             &          & MS$_1$          &                &                 &         &  \\
9-15 &  {\tt UNJS}          &  7.5    & MS          & 2             &          & MS$_1$,LWD$_4$   &                &                 &         &  \\
9-16 &  {\tt UNJS}          &  7.5    & MS          & 1             &          & MS$_1$,LWD$_2$,LWD$_4$   &         &                 &         &  \\
\hline
9-17 &  {\tt UNJS}          &  8.0    & MS          & 3             &          & MS$_1$         &                 &                 &         &  \\
9-18 &  {\tt UNJS}          &  8.0    & MS          & 3             &          &                &   MS$_{1-2}$     &                 &         &  \\
9-19 &  {\tt UNJS}          &  8.0    & MS          & 3             &          &                &   MS$_{1-2}$     &                 &         &  \\
9-20 &  {\tt UNJS}          &  8.0    & MS          & 1             &  MS$_2$  & MS$_1$,MS$_4$   &                 &                 &         &  \\
\hline
9-21 &  {\tt UNJS}          &  8.5    & EWD           & 3             &          & EWD$_1$          &              &                 &   (0/1)    &  \\
9-22 &  {\tt UNJS}          &  8.5    & EWD           & 3             &          & EWD$_2$          &              &                 &   (8/0)    &  \\
9-23 &  {\tt UNJS}          &  8.5    & EWD           & 3             &          & EWD$_1$          &              &                 &   (8/1)    &  \\
9-24 &  {\tt UNJS}          &  8.5    & EWD           & 3             &          & EWD$_2$          &              &                 &   (6/0)    &  \\
\hline
9-25 &  {\tt UNJS}          &  9.0    & EWD          & 3             &          & EWD$_2$          &              &                 &   (2/2)    &  \\
9-26 &  {\tt UNJS}          &  9.0    & EWD           & 3             &          & EWD$_1$          &              &                 &   (4/1)    &  \\
9-27 &  {\tt UNJS}          &  9.0    & EWD           & 3             &          & EWD$_1$          &              &                 &   (0/2)    &  \\
9-28 &  {\tt UNJS}          &  9.0    & EWD          & 3             &          & EWD$_1$          &              &                 &   (7/0)    &  \\
\hline
9-29 &  {\tt UNJS}          &  9.5    &             & 4             &          &                &              &                 &   (0/0)    &  \\
9-30 &  {\tt UNJS}          &  9.5    &             & 4             &          &                &              &                 &   (0/0)    &  \\
9-31 &  {\tt UNJS}          &  9.5    &             & 4             &          &                &              &                 &   (0/0)    &  \\
9-32 &  {\tt UNJS}          &  9.5    &             & 4             &          &                &              &                 &   (0/0)    &  \\
\hline
9-33 &  {\tt \=U\=N\=J\=S}  &  10.0   & MS          & 2             &          &                & MS$_{2-4}$,MS$_{3-4}$    &       &       &  \\
9-34 &  {\tt \=U\=N\=J\=S}  &  10.0   & MS          & 3             &          &                & GB$_{2-3}$    &                 &       &  \\
9-35 &  {\tt \=U\=N\=J\=S}  &  10.0   & MS          & 3             &          &                & MS$_{2-3}$    &                 &       &  \\
9-36 &  {\tt \=U\=N\=J\=S}  &  10.0   & MS          & 4             &          &                &              &                 &       & a \\
\hline
9-37 &  {\tt \=U\=N\=J\=S}  &  11.0   & MS          & 3             &          &                & MS$_{1-3}$    &                 &       &  \\
9-38 &  {\tt \=U\=N\=J\=S}  &  11.0   & MS          & 4             &          &                &              &                 &       &  \\
9-39 &  {\tt \=U\=N\=J\=S}  &  11.0   & MS          & 4             &          &                &              &                 &       &  \\
9-40 &  {\tt \=U\=N\=J\=S}  &  11.0   & MS          & 4             &          &                &              &                 &       &  \\
\hline
9-41 &  {\tt \=U\=N\=J\=S}  &  12.0   & MS          & 3             &          &                & GB$_{1-2}$    &                 &       &  \\
9-42 &  {\tt \=U\=N\=J\=S}  &  12.0   & EWD           & 3             &          &                & LWD$_{3-4}$    &                 &       &  \\
9-43 &  {\tt \=U\=N\=J\=S}  &  12.0   & LWD          & 4             &          &                &              &                 &       &  \\
9-44 &  {\tt \=U\=N\=J\=S}  &  12.0   & EWD           & 4             &          &                &              &                 &       &  \\
\hline
\end{tabular}
\end{minipage}
\end{table*}

\begin{table*}
 \centering
 \begin{minipage}{180mm}
 \centering
  \caption{Summary of results for {\tt NUJS} and {\tt \=N\=U\=J\=S}.  See Sec. \ref{tabcol} for a full description of the columns or Table 
          \ref{JUUU} for a summary. \newline
          MS = main sequence, GB = giant branch, EWD = white dwarf with 0-100 Myr cooling, LWD = white dwarf beyond 100 Myr cooling.\newline  
          Notes:\newline $^{\rm a}$Subsequent evolution may have been affected by tides on the GB phase.\newline
              }
          \label{NUJS}
  \begin{tabular}{@{}ccccccccccc@{}}
 \hline
Sim \#  &  Setup            & $\beta$ & Unpack    & \# Surv & Engulf & Eject           & Collision     & $< R_{\rm max}$? &  TPs Eng & Notes   \\
 \hline
10-1  &  {\tt NUJS}          &  7.0    & MS          & 3        &          &                    &   MS$_{2-3}$    &                 &         &  \\
10-2  &  {\tt NUJS}          &  7.0    & MS          & 2        &          &  MS$_{1}$,MS$_{2}$  &                &                 &         &  \\
10-3  &  {\tt NUJS}          &  7.0    & MS          & 3        &          &  MS$_{2}$           &               &                 &         &  \\
10-4  &  {\tt NUJS}          &  7.0    & MS          & 3        &          &  MS$_{1}$           &               &                 &         &  \\
\hline
10-5  &  {\tt NUJS}          &  7.5    & MS          & 3        &          &  MS$_{2}$           &               &                 &         &  \\
10-6  &  {\tt NUJS}          &  7.5    & MS          & 2        &  EWD$_2$   &  MS$_{1}$           &               &                 &         &  \\
10-7  &  {\tt NUJS}          &  7.5    & MS          & 2        &  MS$_2$  &  MS$_{1}$           &               &                 &         &  \\
10-8  &  {\tt NUJS}          &  7.5    & MS          & 2        &          &  MS$_{2}$,LWD$_1$    &               &                 &         &  \\
\hline
10-9  &  {\tt NUJS}          &  8.0    & EWD           & 2        &          &  EWD$_{1}$,EWD$_2$     &               &                 &         &  \\
10-10 &  {\tt NUJS}          &  8.0    & MS           & 3        &          &  MS$_{2}$          &               &                 &         &  \\
10-11 &  {\tt NUJS}          &  8.0    & MS          & 3        &          &  MS$_{2}$          &               &                 &         & a \\
10-12 &  {\tt NUJS}          &  8.0    & MS          & 3        &          &  MS$_{2}$          &               &                 &         &  \\
\hline
10-13 &  {\tt NUJS}          &  8.5    & EWD           & 3        &          &  EWD$_{2}$          &               &                 &   (6/0)  &  \\
10-14 &  {\tt NUJS}          &  8.5    & EWD           & 3        &          &  EWD$_{1}$          &               &                 &   (5/0)  &  \\
10-15 &  {\tt NUJS}          &  8.5    & EWD           & 3        &          &  EWD$_{2}$          &               &                 &   (1/3)  &  \\
10-16 &  {\tt NUJS}          &  8.5    & EWD           & 3        &          &  EWD$_{2}$          &               &                 &   (2/1)  &  \\
\hline
10-17 &  {\tt NUJS}          &  9.0    & MS          & 3        &          &  MS$_{2}$         &               &                 &   (0/0)  &  \\
10-18 &  {\tt NUJS}          &  9.0    & EWD           & 3        &          &  EWD$_{2}$          &               &                 &   (0/4)  &  \\
10-19 &  {\tt NUJS}          &  9.0    & EWD           & 2        &          &  EWD$_1$,EWD$_{2}$    &               &                 &   (5/0)  &  \\
10-20 &  {\tt NUJS}          &  9.0    & EWD           & 1        &          &  EWD$_1$,EWD$_{2}$,LWD$_4$ &           &                 &   (4/1)  &  \\
\hline
10-21 &  {\tt NUJS}          &  9.5    &             & 4        &          &                  &               &                 &   (0/0)  &  \\
10-22 &  {\tt NUJS}          &  9.5    &             & 4        &          &                  &               &                 &   (0/0)  &  \\
10-23 &  {\tt NUJS}          &  9.5    &             & 4        &          &                  &               &                 &   (0/0)  &  \\
10-24 &  {\tt NUJS}          &  9.5    &             & 4        &          &                  &               &                 &   (0/0)  &  \\
\hline
10-25 &  {\tt \=N\=U\=J\=S}  &  10.0   & EWD           & 3        &          &                  &   EWD$_{2-3}$    &                 &          &  \\
10-26 &  {\tt \=N\=U\=J\=S}  &  10.0   & MS          & 3        &          &                  &  MS$_{1-2}$    &                 &          &  \\
10-27 &  {\tt \=N\=U\=J\=S}  &  10.0   & MS          & 4        &          &                  &               &                 &          &  \\
10-28 &  {\tt \=N\=U\=J\=S}  &  10.0   & MS          & 3        &          &                  &  MS$_{3-4}$    &                 &          &  \\
\hline
10-29 &  {\tt \=N\=U\=J\=S}  &  11.0   & MS          & 3        &          &                  &  MS$_{1-4}$    &                 &          &  \\
10-30 &  {\tt \=N\=U\=J\=S}  &  11.0   & MS          & 3        &          &                  &  MS$_{3-4}$    &                 &          &  \\
10-31 &  {\tt \=N\=U\=J\=S}  &  11.0   & MS          & 3        &          &                  &  MS$_{1-3}$    &                 &          & a \\
10-32 &  {\tt \=N\=U\=J\=S}  &  11.0   & MS          & 3        &          &                  &  MS$_{2-3}$    &                 &          &  \\
\hline
10-33 &  {\tt \=N\=U\=J\=S}  &  12.0   & MS          & 3        &          &                  &  LWD$_{3-4}$    &                 &          &  \\
10-34 &  {\tt \=N\=U\=J\=S}  &  12.0   & MS          & 3        &          &                  &  MS$_{2-3}$    &                 &          &  \\
10-35 &  {\tt \=N\=U\=J\=S}  &  12.0   & MS          & 3        &          &                  &  MS$_{2-4}$    &                 &          &  \\
10-36 &  {\tt \=N\=U\=J\=S}  &  12.0   & MS          & 3        &          &                  &  MS$_{3-4}$    &                 &          &  \\
\hline
\end{tabular}
\end{minipage}
\end{table*}

\begin{table*}
 \centering
 \begin{minipage}{180mm}
 \centering
  \caption{Summary of results for {\tt JUNS} and {\tt \=J\=U\=N\=S}.  See Sec. \ref{tabcol} for a full description of the columns or Table 
          \ref{JUUU} for a summary. \newline
          MS = main sequence, GB = giant branch, EWD = white dwarf with 0-100 Myr cooling, LWD = white dwarf beyond 100 Myr cooling.\newline
          Notes:\newline $^{\rm a}$Subsequent evolution may have been affected by tides on the GB phase.\newline 
                  $^{\rm d}$A test particle which survived the entire integration acheived a minimum pericentre of 0.039 au at a WD cooling age of 10.091 Gyr.\newline
                  $^{\rm z}$Unpacking, ejections and engulfments all occur for WD cooling ages exceeding 10 Gyr.}
          \label{JUNS}
  \begin{tabular}{@{}ccccccccccc@{}}
 \hline
Sim \#  &  Setup            & $\beta$ & Unpack    & \# Surv & Engulf & Eject         & Collision     & $< R_{\rm max}$? & TPs Eng  & Notes   \\
 \hline
11-1  &  {\tt JUNS}         &  7.0    & MS          & 2        &          &  MS$_2$,MS$_3$      &           &                 &         &  \\
11-2  &  {\tt JUNS}         &  7.0    & MS          & 2        &          &  MS$_2$,MS$_3$      &           &                 &         &  \\
11-3  &  {\tt JUNS}         &  7.0    & MS          & 1        &          &  MS$_3$,EWD$_4$       & MS$_{1-2}$ &                 &         &  \\
11-4  &  {\tt JUNS}         &  7.0    & MS          & 2        &          &  MS$_2$,MS$_3$      &           &                 &         &  \\
\hline
11-5  &  {\tt JUNS}         &  7.5    & MS          & 2        &          &  MS$_2$,MS$_3$      &           &                 &         &  \\
11-6  &  {\tt JUNS}         &  7.5    & MS          & 2        &          &  MS$_2$,MS$_3$      &           &                 &         &  \\
11-7  &  {\tt JUNS}         &  7.5    & MS          & 2        &   MS$_3$ &  MS$_2$             &           &                 &         &  \\
11-8  &  {\tt JUNS}         &  7.5    & MS          & 2        &          &  MS$_2$,MS$_3$      &           &                 &         &  \\
\hline
11-9  &  {\tt JUNS}         &  8.0    & EWD           & 2        &          &  EWD$_3$,LWD$_2$       &           &                 &         &  \\
11-10 &  {\tt JUNS}         &  8.0    & MS          & 2        &          &  MS$_2$,MS$_3$      &           &                 &         &  \\
11-11 &  {\tt JUNS}         &  8.0    & MS          & 2        &          &  MS$_2$,EWD$_3$       &           &                 &         &  \\
11-12 &  {\tt JUNS}         &  8.0    & EWD          & 2        &          &  EWD$_2$,LWD$_3$       &           &                 &         &  \\
\hline
11-13 &  {\tt JUNS}         &  8.5    & EWD           & 2        &          &  EWD$_3$              & EWD$_{1-2}$  &                 &   (0/0) &  \\
11-14 &  {\tt JUNS}         &  8.5    & EWD           & 2        &          &  EWD$_2$,LWD$_3$       &           &                 &   (4/0) &  \\
11-15 &  {\tt JUNS}         &  8.5    & EWD           & 2        &          &  EWD$_2$,EWD$_3$        &           &                 &   (2/0) &  \\
11-16 &  {\tt JUNS}         &  8.5    & EWD           & 2        &          &  EWD$_2$,LWD$_3$       &           &                 &   (1/0) &  \\
\hline
11-17 &  {\tt JUNS}         &  9.0    & LWD          & 2        &          &  LWD$_2$,LWD$_3$      &           &                 &   (0/3) &  \\
11-18 &  {\tt JUNS}         &  9.0    & LWD          & 3        &          &  LWD$_2$             &           &                 &   (0/4) &  \\
11-19 &  {\tt JUNS}         &  9.0    &             & 4        &          &                     &           &                 &   (0/1) &  \\
11-20 &  {\tt JUNS}         &  9.0    & LWD          & 2        &          &  LWD$_2$,LWD$_3$      &           &                 &   (0/1) &  \\
\hline
11-21 &  {\tt JUNS}         &  9.5    & LWD          & 3        &          &                    & LWD$_{1-3}$ &                 &   (0/2) &  \\
11-22 &  {\tt JUNS}         &  9.5    & LWD          & 2        &          &  LWD$_2$,LWD$_3$      &           &                 &   (0/0) &  \\
11-23 &  {\tt JUNS}         &  9.5    & LWD          & 2        &          &  LWD$_2$,LWD$_3$      &           &                 &   (0/4) & d,z \\
11-24 &  {\tt JUNS}         &  9.5    &             & 4        &          &                     &           &                 &   (0/0) &  \\
\hline
11-25 &  {\tt \=J\=U\=N\=S} &  10.0   & MS          & 4        &          &                     &           &                 &         & a \\
11-26 &  {\tt \=J\=U\=N\=S} &  10.0   & MS          & 4        &          &                     &           &   \#2$_{0.956}$  &         &   \\
11-27 &  {\tt \=J\=U\=N\=S} &  10.0   & MS          & 3        &          &                     & GB$_{1-2}$ &                 &         &   \\
11-28 &  {\tt \=J\=U\=N\=S} &  10.0   & MS          & 4        &          &                     &           &                 &         & a \\
\hline
11-29 &  {\tt \=J\=U\=N\=S} &  11.0   & MS          & 4        &          &                     &           &                 &         &   \\
11-30 &  {\tt \=J\=U\=N\=S} &  11.0   & MS          & 3        &          &                     & MS$_{3-4}$ &                 &         &   \\
11-31 &  {\tt \=J\=U\=N\=S} &  11.0   & MS          & 2        &          &                     & MS$_{1-3}$,LWD$_{1-4}$ &       &         &   \\
11-32 &  {\tt \=J\=U\=N\=S} &  11.0   & MS          & 3        &          &                     & GB$_{1-4}$ &                 &         &   \\
\hline
11-33 &  {\tt \=J\=U\=N\=S} &  12.0   & EWD           & 4        &          &                     &           &                 &         &   \\
11-34 &  {\tt \=J\=U\=N\=S} &  12.0   & MS          & 2        &          &                     & MS$_{1-2}$,LWD$_{3-4}$  &      &         &   \\
11-35 &  {\tt \=J\=U\=N\=S} &  12.0   & MS          & 4        &          &                     &           &   \#3$_{1.764}$  &         &   \\
11-36 &  {\tt \=J\=U\=N\=S} &  12.0   & EWD           & 3        &          &                     & LWD$_{1-4}$          &        &         &   \\
\hline
\end{tabular}
\end{minipage}
\end{table*}

\begin{table*}
 \centering
 \begin{minipage}{180mm}
 \centering
  \caption{Summary of results for initially alternating Jupiters and Saturns.  See Sec. \ref{tabcol} for a full description of the columns or Table 
          \ref{JUUU} for a summary. \newline
          MS = main sequence, GB = giant branch, EWD = white dwarf with 0-100 Myr cooling, LWD = white dwarf beyond 100 Myr cooling.\newline
          Notes: \newline
                   $^{\rm c}$Subsequent evolution may have been affected by tides on the WD phase.   \newline
                   $^{\rm k}$The simulation ran for just 2.534 Gyr due to the tight orbit of the fourth planet along the white dwarf phase.\newline
                   $^{\rm l}$The simulation ran for just 4.644 Gyr due to the tight orbit of the sixth planet along the white dwarf phase.\newline
                   $^{\rm m}$The simulation ran for just 2.994 Gyr due to the tight orbit of the fifth planet along the white dwarf phase.\newline
                   $^{\rm n}$The simulation ran for just 3.823 Gyr due to the tight orbit of the fifth planet along the white dwarf phase.\newline
                   $^{\rm o}$The simulation ran for just 8.433 Gyr due to the tight orbit of the third planet along the white dwarf phase.\newline
                   $^{\rm p}$The simulation ran for just 7.105 Gyr due to the tight orbit of the first planet along the white dwarf phase.
                   }
          \label{JSJS}
  \begin{tabular}{@{}ccccccccccc@{}}
 \hline
Sim \#  &  Setup            & $\beta$ & Unpack    & \# Surv & Engulf & Eject           & Collision     & $< R_{\rm max}$?  & Notes   \\
 \hline
12-1  &  {\tt JSJS}         &  6.0    & EWD          & 2        &          &  EWD$_2$,EWD$_4$      &           &                   &  \\
12-2  &  {\tt JSJS}         &  6.0    & EWD          & 2        &          &  EWD$_2$,EWD$_4$      &           &                   &  \\
12-3  &  {\tt JSJS}         &  6.0    & EWD           & 2        &   EWD$_4$  &  EWD$_2$            &           &                   &  \\
12-4  &  {\tt JSJS}         &  6.0    & EWD          & 2        &          &  EWD$_2$,EWD$_4$      &           &                   &  \\
\hline
12-5 &  {\tt JSJS}          &  7.0    & EWD          & 2        &          &  EWD$_2$,EWD$_4$        &          &                   &  \\
12-6 &  {\tt JSJS}          &  7.0    & EWD         & 2        &          &  EWD$_2$,EWD$_4$,LWD$_3$ &          &      \#1$_{0.428}$  &  \\
12-7 &  {\tt JSJS}          &  7.0    & EWD          & 2        &          &  EWD$_2$,EWD$_4$    &          &             &  \\
12-8 &  {\tt JSJS}          &  7.0    & EWD          & 3        &          &  EWD$_2$        &          &     \#4$_{0.218}$     & k \\
\hline
12-9  &  {\tt JSJSJS}       &  6.0    & EWD          & 3        &  EWD$_4$   & EWD$_2$,EWD$_6$      &           &                   &  \\
12-10  &  {\tt JSJSJS}      &  6.0    & MS         & 2        &  MS$_2$,GB$_{1}$   &  MS$_4$,MS$_6$      &           &           &  \\
12-11  &  {\tt JSJSJS}      &  6.0    & EWD          & 3        &  EWD$_2$   &  EWD$_3$,LWD$_4$      &           &   \#6$_{0.917}$     & l \\
\hline
12-12  &  {\tt JSJSJS}      &  7.0    & EWD          & 2        &     &  EWD$_1$,EWD$_2$,EWD$_4$,LWD$_6$     &           &              &  \\
12-13  &  {\tt JSJSJS}      &  7.0    & EWD          & 2        &  EWD$_3$   &  EWD$_2$,EWD$_4$,EWD$_6$     &           &              &  \\
12-14  &  {\tt JSJSJS}      &  7.0    & EWD          & 1        &  LWD$_3$   &  EWD$_2$,EWD$_4$,EWD$_6$,LWD$_5$   &           &              &  \\
12-15  &  {\tt JSJSJS}      &  7.0    & EWD          & 3        &  EWD$_6$   &  EWD$_2$,EWD$_4$  &           &              &  \\
\hline
12-16  &  {\tt JSJSJSJS}    &  6.0    & EWD          & 3        &  EWD$_1$  &  EWD$_4$,EWD$_6$,EWD$_8$,LWD$_3$      &   &   \#5$_{0.669}$     & m \\
12-17  &  {\tt JSJSJSJS}    &  6.0    & EWD          & 3        &  EWD$_7$  &  EWD$_3$,EWD$_4$,EWD$_6$,EWD$_8$      &   &   \#1$_{0.473}$, \#5$_{0.354}$   & n \\
12-18  &  {\tt JSJSJSJS}    &  6.0    & EWD          & 2        &   &  EWD$_1$,EWD$_2$,EWD$_5$,EWD$_6$,LWD$_3$,LWD$_4$    &   &   \#7$_{0.381}$   &  \\
12-19  &  {\tt JSJSJSJS}    &  6.0    & MS         & 2        &  MS$_6$,GB$_3$ &  MS$_2$,MS$_4$,MS$_7$,MS$_8$    &   &     &  \\
\hline
12-20  &  {\tt JSJSJSJS}    &  7.0    & EWD          & 3        &   &  EWD$_1$,EWD$_4$,EWD$_8$,LWD$_7$   & EWD$_{1-2}$  &     &  \\
12-21  &  {\tt JSJSJSJS}    &  7.0    & EWD          & 3        &   &  EWD$_4$,EWD$_5$,EWD$_6$,EWD$_7$,EWD$_8$   &   &  \#2$_{1.566}$, \#3$_{0.034}$   & c,o \\
12-22  &  {\tt JSJSJSJS}    &  7.0    & EWD          & 2        &   &  EWD$_2$,EWD$_3$,EWD$_4$,EWD$_8$,LWD$_5$,LWD$_6$   &   &  \#1$_{0.171}$    & c \\
12-23  &  {\tt JSJSJSJS}    &  7.0    & EWD          & 3        & EWD$_6$  &  EWD$_2$,EWD$_4$,EWD$_5$,LWD$_8$   &   &  \#1$_{1.586}$    & p \\
\hline
\end{tabular}
\end{minipage}
\end{table*}

\begin{table*}
 \centering
 \begin{minipage}{180mm}
 \centering
  \caption{Summary of results for initially alternating Uranuses and Neptunes.  See Sec. \ref{tabcol} for a full description of the columns or Table 
          \ref{JUUU} for a summary. \newline
          MS = main sequence, GB = giant branch, EWD = white dwarf with 0-100 Myr cooling, LWD = white dwarf beyond 100 Myr cooling.\newline
          Notes: \newline
                   $^{\rm b}$Subsequent evolution may have been affected by tides on the MS phase.\newline
                   $^{\rm c}$Subsequent evolution may have been affected by tides on the WD phase.\newline
                   $^{\rm q}$The simulation ran for just 4.831 Gyr due to the tight orbit of the first planet along the white dwarf phase.\newline
                   $^{\rm r}$The simulation ran for just 2.196 Gyr due to the very tight orbits of the first and sixth planets along the white dwarf phase.\newline
                   $^{\rm s}$The simulation ran for just 7.368 Gyr due to the tight orbit of the fourth planet along the white dwarf phase.\newline
                   $^{\rm t}$The simulation ran for just 2.564 Gyr due to the tight orbit of the fourth planet along the white dwarf phase.\newline
                   $^{\rm u}$The simulation ran for just 2.078 Gyr due to the tight orbit of the first planet along the white dwarf phase.\newline
                   $^{\rm v}$The simulation ran for just 1.926 Gyr due to the tight orbit of the eighth planet along the white dwarf phase.
                   }
          \label{UNUN}
  \begin{tabular}{@{}ccccccccccc@{}}
 \hline
Sim \#  &  Setup            & $\beta$ & Unpack    & \# Surv & Engulf & Eject          & Collision     & $< R_{\rm max}$?  & Notes   \\
 \hline
13-1  &  {\tt UNUN}         &  7.0    & EWD          & 3        &    EWD$_2$      &        &           &     \#1$_{1.769}$, \#3$_{0.00731}$      & c \\
13-2  &  {\tt UNUN}         &  7.0    & EWD          & 2        &               &  EWD$_2$      &   EWD$_{1-3}$    &      \#4$_{0.608}$      &  \\
13-3  &  {\tt UNUN}         &  7.0    & EWD          & 3        &    EWD$_2$      &           &          &      \#1$_{0.797}$      &  \\
13-4  &  {\tt UNUN}         &  7.0    & GB         & 3        &   GB$_2$      &           &          &            &  \\
\hline
13-5  &  {\tt UNUN}         &  9.0    & LWD         & 2        &               &  LWD$_2$      &   LWD$_{1-3}$    &      \#1$_{0.891}$      &  \\
13-6  &  {\tt UNUN}         &  9.0    & LWD         & 2        &               &    &   LWD$_{1-2}$,LWD$_{1-3}$    &            &  \\
13-7  &  {\tt UNUN}         &  9.0    & LWD         & 2        &   LWD$_1$      &  LWD$_4$  &      &     \#2$_{0.00935}$       & c \\
13-8  &  {\tt UNUN}         &  9.0    & LWD         & 3        &   LWD$_3$      &     &      &             &  \\
\hline
13-9  &  {\tt UNUNUN}       &  7.0    & MS         & 1        &   MS$_1$,GB$_6$   &  MS$_4$,MS$_5$    &  MS$_{2-3}$    &            & b \\
13-10 &  {\tt UNUNUN}       &  7.0    & MS         & 3        &   MS$_5$,GB$_6$   &    &  MS$_{2-4}$    &      \#2$_{0.906}$     &  \\
13-11 &  {\tt UNUNUN}       &  7.0    & MS         & 3        &   MS$_1$   &    &  MS$_{1-3}$,MS$_{2-4}$    &       &  \\
\hline
13-12 &  {\tt UNUNUN}       &  9.0    & EWD          & 4        &   LWD$_2$   &  LWD$_6$   &      &  \#1$_{0.784}$, \#3$_{0.0435}$     & c,q \\
13-13 &  {\tt UNUNUN}       &  9.0    & EWD          & 5        &   LWD$_2$   &     &      &  \#1$_{0.0424}$, \#5$_{0.667}$, \#6$_{0.211}$   & c,r \\
13-14 &  {\tt UNUNUN}       &  9.0    & EWD          & 4        &   LWD$_2$   &  LWD$_1$    &      &  \#3$_{0.238}$, \#4$_{0.0170}$, \#5$_{0.0187}$   & c,s \\
\hline
13-15 &  {\tt UNUNUNUN}     &  9.0    & EWD          & 5        &   EWD$_3$   &  LWD$_1$,LWD$_2$   &      &  \#4$_{0.0889}$, \#8$_{0.153}$   & c,t \\
13-16 &  {\tt UNUNUNUN}     &  9.0    & EWD          & 7        &   LWD$_8$   &     &      &  \#1$_{0.0121}$, \#7$_{0.133}$   & c,u \\
13-17 &  {\tt UNUNUNUN}     &  9.0    & EWD          & 7        &   EWD$_1$   &     &      &  \#3$_{0.213}$, \#8$_{0.293}$   & v \\
\hline
\end{tabular}
\end{minipage}
\end{table*}

\label{lastpage}

\begin{thebibliography}{99}

\bibitem[Adams et al.(2013)]{adaetal2013} Adams, F.~C., Anderson, K.~R., \& Bloch, A.~M.\ 2013, MNRAS, 432, 438 

\bibitem[Andrews et al.(2013)]{andetal2013} Andrews, S.~M., Rosenfeld, K.~A., Kraus, A.~L., \& Wilner, D.~J.\ 2013, ApJ, 771, 129 

\bibitem[Barclay et al.(2015)]{baretal2015} Barclay, T., Quintana, E.~V., Adams, F.~C., et al.\ 2015, ApJ, 809, 7 

\bibitem[Bear \& Soker(2013)]{beasok2013} Bear, E., \& Soker, N.\ 2013, New Astronomy, 19, 56

\bibitem[Bergfors et al.(2014)]{beretal2014} Bergfors, C., Farihi, 
J., Dufour, P., \& Rocchetto, M.\ 2014, MNRAS, 444, 2147 

\bibitem[Bonsor et al.(2011)]{bonetal2011} Bonsor, A., Mustill, A.~J., \& Wyatt, M.~C.\ 2011, MNRAS, 414, 930 

\bibitem[Bonsor \& Veras(2015)]{bonver2015} Bonsor, A., \& Veras, D.\ 2015, MNRAS, 454, 53 

\bibitem[Campante et al.(2015)]{cametal2015} Campante, T.~L., Barclay, T., Swift, J.~J., et al.\ 2015, ApJ, 799, 170 

\bibitem[Casewell et al.(2009)]{casetal2009} Casewell, S.~L., Dobbie, P.~D., Napiwotzki, R., et al.\ 2009, MNRAS, 395, 1795

\bibitem[Cassan et al.(2012)]{casetal2012} Cassan, A., Kubas, D., Beaulieu, J.-P., et al.\ 2012, Nature, 481, 167

\bibitem[Catal{\'a}n et al.(2008)]{catetal2008} Catal{\'a}n, S., Isern, J., Garc{\'{\i}}a-Berro, E., \& Ribas, I.\ 2008, 
MNRAS, 387, 1693 

\bibitem[Chambers et al.(1996)]{chaetal1996} Chambers, J.~E., Wetherill, G.~W., \& Boss, A.~P.\ 1996, Icarus, 119, 261 

\bibitem[Chambers(1999)]{chambers1999} Chambers, J.~E.\ 1999, MNRAS, 304, 793 

\bibitem[Croll et al.(2016)]{croetal2016} Croll, B., Dalba, P.~A., Vanderburg, A., et al.\ 2016, Submitted to ApJL, 
arXiv:1510.06434 

\bibitem[Davies et al.(2014)]{davetal2014} Davies, M.~B., Adams, F.~C., Armitage, P., et al.\ 2014, Protostars and Planets VI, 787 

\bibitem[Debes \& Sigurdsson(2002)]{debsig2002} Debes, J.~H., \& Sigurdsson, S.\ 2002, ApJ, 572, 556 

\bibitem[Debes et al.(2012)]{debetal2012} Debes, J.~H., Walsh, K.~J., \& Stark, C.\ 2012, ApJ, 747, 148

\bibitem[Dong et al.(2010)]{donetal2010} Dong, R., Wang, Y., Lin, D.~N.~C., \& Liu, X.-W.\ 2010, ApJ, 715, 1036

\bibitem[Dufour et al.(2007)]{dufetal2007} Dufour, P., Bergeron, P., Liebert, J., et al.\ 2007, ApJ, 663, 1291

\bibitem[Duncan \& Lissauer(1998)]{dunlis1998} Duncan, M.~J., \& Lissauer, J.~J.\ 1998, Icarus, 134, 303 

\bibitem[Faber \& Quillen(2007)]{fabqui2007} Faber, P., \& Quillen, A.~C.\ 2007, MNRAS, 382, 1823 

\bibitem[Falcon et al.(2010)]{faletal2010} Falcon, R.~E., Winget, D.~E., Montgomery, M.~H., \& Williams, K.~A.\ 2010, ApJ, 712, 585

\bibitem[Farihi et al.(2009)]{faretal2009} Farihi, J., Jura, M., \& Zuckerman, B.\ 2009, ApJ, 694, 805 

\bibitem[Farihi et al.(2010)]{faretal2010} Farihi, J., Barstow, M.~A., Redfield, S., Dufour, P., \& Hambly, N.~C.\ 2010, MNRAS, 404, 2123 

\bibitem[Farihi(2016)]{farihi2016} Farihi, J. 2016 Submitted, New Astronomy Reviews.


\bibitem[Frewen \& Hansen(2014)]{frehan2014} Frewen, S.~F.~N., \& Hansen, B.~M.~S.\ 2014, MNRAS, 439, 2442 

\bibitem[G\"{a}nsicke et al.(2016)]{gaeetal2016} G\"{a}nsicke, B.~T., Aungwerojwit, A., Marsh, T.~R. et al.\ 2016, In Press ApJL, arXiv:1512.09150

\bibitem[Gentile Fusillo et al.(2015)]{genetal2015} Gentile Fusillo, N.~P., G{\"a}nsicke, B.~T., \& Greiss, S.\ 2015, MNRAS, 448, 2260 

\bibitem[Girven et al.(2011)]{giretal2011} Girven, J., G{\"a}nsicke, B.~T., Steeghs, D., \& Koester, D.\ 2011, MNRAS, 417, 1210 

\bibitem[Girven et al.(2012)]{giretal2012} Girven, J., Brinkworth, C.~S., Farihi, J., et al.\ 2012, ApJ, 749, 154 

\bibitem[Gomes et al.(2005)]{gometal2005} Gomes, R., Levison, H.~F., Tsiganis, K., \& Morbidelli, A.\ 2005, Nature, 435, 466

\bibitem[Graham et al.(1990)]{graetal1990} Graham, J.~R., Matthews, K., Neugebauer, G., \& Soifer, B.~T.\ 1990, ApJ, 357, 216 

\bibitem[Guillochon et al.(2011)]{guietal2011} Guillochon, J., Ramirez-Ruiz, E., \& Lin, D.\ 2011, ApJ, 732, 74

\bibitem[Henning \& Hurford(2014)]{henhur2014} Henning, W.~G., \& Hurford, T.\ 2014, ApJ, 789, 30 


\bibitem[Hurley et al.(2000)]{huretal2000} Hurley, J.~R., Pols, O.~R., \& Tout, C.~A.\ 2000, MNRAS, 315, 543

\bibitem[Izidoro et al.(2015)]{izietal2015} Izidoro, A., Raymond, S.~N., Morbidelli, A., \& Winter, O.~C.\ 2015, MNRAS, 453, 3619

\bibitem[Jura(2003)]{jura2003} Jura, M.\ 2003, ApJL, 584, L91 

\bibitem[Kalirai et al.(2008)]{kaletal2008} Kalirai, J.~S., Hansen, B.~M.~S., Kelson, D.~D., et al.\ 2008, ApJ, 676, 594 

\bibitem[Kennedy \& Kenyon(2008)]{kenken2008} Kennedy, G.~M., \& Kenyon, S.~J.\ 2008, ApJ, 673, 502 

\bibitem[Kepler et al.(2015)]{kepetal2015} Kepler, S.~O., Pelisoli, I., Koester, D., et al.\ 2015, MNRAS, 446, 4078 

\bibitem[Kepler et al.(2016)]{kepetal2016} Kepler, S.~O., Pelisoli, I., Koester, D., et al.\ 2016, MNRAS, 455, 3413

\bibitem[Kleinman et al.(2013)]{kleetal2013} Kleinman, S.~J., Kepler, S.~O., Koester, D., et al.\ 2013, ApJS, 204, 5

\bibitem[Koester et al.(2014)]{koeetal2014} Koester, D., G{\"a}nsicke, B.~T., \& Farihi, J.\ 2014, A\&A, 566, A34 

\bibitem[Levison et al.(1998)]{levetal1998} Levison, H.~F., Lissauer, J.~J., \& Duncan, M.~J.\ 1998, AJ, 116, 1998 

\bibitem[Levison et al.(2011)]{levetal2011} Levison, H.~F., Morbidelli, A., Tsiganis, K., Nesvorn{\'y}, D., \& Gomes, R.\ 2011, AJ, 142, 152 

\bibitem[Liebert et al.(2005)]{lieetal2005} Liebert, J., Bergeron, P., \& Holberg, J.~B.\ 2005, ApJS, 156, 47 

\bibitem[Liu et al.(2013)]{liuetal2013} Liu, S.-F., Guillochon, J., Lin, D.~N.~C., \& Ramirez-Ruiz, E.\ 2013, ApJ, 762, 37 

\bibitem[Luhman et al.(2011)]{luhetal2011} Luhman, K.~L., Burgasser, A.~J., \& Bochanski, J.~J.\ 2011, ApJL, 730, L9 

\bibitem[Marois et al.(2010)]{maretal2010} Marois, C., Zuckerman, B., Konopacky, Q.~M., Macintosh, B., \& Barman, T.\ 2010, Nature, 468, 1080

\bibitem[Marzari \& Weidenschilling(2002)]{marwei2002} Marzari, F., \& Weidenschilling, S.~J.\ 2002, Icarus, 156, 570 

\bibitem[Matsumura et al.(2013)]{matetal2013} Matsumura, S., Ida, S., \& Nagasawa, M.\ 2013, ApJ, 767, 129 

\bibitem[Morbidelli et al.(2005)]{moretal2005} Morbidelli, A., Levison, H.~F., Tsiganis, K., \& Gomes, R.\ 2005, Nature, 435, 462 

\bibitem[Mustill \& Villaver(2012)]{musvil2012} Mustill, A.~J., \& Villaver, E.\ 2012, ApJ, 761, 121 

\bibitem[Mustill et al.(2013)]{musetal2013} Mustill, A.~J., Marshall, J.~P., Villaver, E., et al.\ 2013, MNRAS, 436, 2515 

\bibitem[Mustill et al.(2014)]{musetal2014} Mustill, A.~J., Veras, D., \& Villaver, E.\ 2014, MNRAS, 437, 1404 

\bibitem[Nordhaus \& Spiegel(2013)]{norspi2013} Nordhaus, J., \& Spiegel, D.~S.\ 2013, MNRAS, 432, 500 

\bibitem[O'Brien et al.(2014)]{obretal2014} O'Brien, D.~P., Walsh, K.~J., Morbidelli, A., Raymond, S.~N., 
\& Mandell, A.~M.\ 2014, Icarus, 239, 74 

\bibitem[Paxton et al.(2011)]{paxetal2011} Paxton, B., Bildsten, L., Dotter, A., et al.\ 2011, ApJS, 192, 3 

\bibitem[Paxton et al.(2015)]{paxetal2015} Paxton, B., Marchant, P., Schwab, J., et al.\ 2015, ApJS, 220, 15

\bibitem[Payne et al.(2016)]{payetal2016} Payne, M.~J., Veras, D., Holman, M.~J., et al. \ 2016, MNRAS, 457, 217

\bibitem[Pierens \& Raymond(2011)]{pieray2011} Pierens, A., \& Raymond, S.~N.\ 2011, A\&A, 533, A131 

\bibitem[Portegies Zwart(2013)]{portegieszwart2013} Portegies Zwart, S.\ 2013, MNRAS, 429, L45 

\bibitem[Pu \& Wu(2015)]{puwu2015} Pu, B., \& Wu, Y.\ 2015, ApJ, 807, 44 

\bibitem[Rappaport et al.(2016)]{rapetal2016} Rappaport, S., Gary, B.~L., Kaye, T., et al.\ 2016, Submitted to MNRAS, arXiv:1602.00740 

\bibitem[Rasio \& Ford(1996)]{rasfor1996} Rasio, F.~A., \& Ford, E.~B.\ 1996, Science, 274, 954

\bibitem[Raymond et al.(2011)]{rayetal2011} Raymond, S.~N., Armitage, P.~J., Moro-Mart{\'{\i}}n, A., et al.\ 2011, A\&A, 530, A62 

\bibitem[Raymond et al.(2012)]{rayetal2012} Raymond, S.~N., Armitage, P.~J., Moro-Mart{\'{\i}}n, A., et al.\ 2012, A\&A, 541, A11 


\bibitem[Reffert et al.(2015)]{refetal2015} Reffert, S., Bergmann, C., Quirrenbach, A., Trifonov, T., K\"{u}nstler, A.\ 2015, A\&A, 574, A116 

\bibitem[Safronov \& Zvjagina(1969)]{safzvj1969} Safronov, V.~S., \& Zvjagina, E.~V.\ 1969, Icarus, 10, 109 

\bibitem[Schr{\"o}der \& Connon Smith(2008)]{schcon2008} Schr{\"o}der, K.-P., \& Connon Smith, R.\ 2008, MNRAS, 386, 155 

\bibitem[Schr{\"o}der \& Cuntz(2005)]{schcun2005} Schr{\"o}der, K.-P., \& Cuntz, M.\ 2005, ApJL, 630, L73 

\bibitem[Sigurdsson et al.(2003)]{sigetal2003} Sigurdsson, S., Richer, H.~B., Hansen, B.~M., Stairs, I.~H., 
\& Thorsett, S.~E.\ 2003, Science, 301, 193 

\bibitem[Smith \& Lissauer(2009)]{smilis2009} Smith, A.~W., \& Lissauer, J.~J.\ 2009, Icarus, 201, 381 

\bibitem[Smullen et al.(2016)]{smuetal2016} Smullen, R.~A., Kratter, K.~M., Shannon, A.\ 2016, Submitted to ApJ

\bibitem[Steele et al.(2011)]{steetal2011} Steele, P.~R., Burleigh, M.~R., Dobbie, P.~D., et al.\ 2011, MNRAS, 416, 2768

\bibitem[Sumi et al.(2011)]{sumetal2011} Sumi, T., Kamiya, K., Bennett, D.~P., et al.\ 2011, Nature, 473, 349 

\bibitem[Sutherland \& Fabrycky(2015)]{sutfab2015} Sutherland, A.~P., \& Fabrycky, D.~C.\ 2015, Submitted to ApJ, arXiv:1511.03274 

\bibitem[Tremblay et al.(2013)]{treetal2013} Tremblay, P.-E., Ludwig, H.-G., Steffen, M., \& Freytag, B.\ 2013, A\&A, 559, A104 

\bibitem[Tsiganis et al.(2005)]{tsietal2005} Tsiganis, K., Gomes, R., Morbidelli, A., \& Levison, H.~F.\ 2005, Nature, 435, 459 

\bibitem[Vanderburg et al.(2015)]{vanetal2015} Vanderburg, A., Johnson, J.~A., Rappaport, S., et al.\ 2015, Nature, 526, 546 

\bibitem[Veras \& Armitage(2005)]{verarm2005} Veras, D., \& Armitage, P.~J.\ 2005, ApJL, 620, L111 

\bibitem[Veras \& Armitage(2006)]{verarm2006} Veras, D., \& Armitage, P.~J.\ 2006, ApJ, 645, 1509 

\bibitem[Veras \& Ford(2010)]{verfor2010} Veras, D., \& Ford, E.~B.\ 2010, ApJ, 715, 803 

\bibitem[Veras et al.(2011)]{veretal2011} Veras, D., Wyatt, M.~C., Mustill, A.~J., Bonsor, A., \& Eldridge, J.~J.\ 2011, MNRAS, 417, 2104 

\bibitem[Veras \& Moeckel(2012)]{vermoe2012} Veras, D., \& Moeckel, N.\ 2012, MNRAS, 425, 680 

\bibitem[Veras \& Raymond(2012)]{verray2012} Veras, D., \& Raymond, S.~N.\ 2012, MNRAS, 421, L117 

\bibitem[Veras \& Tout(2012)]{vertou2012} Veras, D., \& Tout, C.~A.\ 2012, MNRAS, 422, 1648 

\bibitem[Veras \& Evans(2013)]{vereva2013} Veras, D., \& Evans, N.~W.\ 2013, MNRAS, 430, 403 

\bibitem[Veras et al.(2013a)]{veretal2013a} Veras, D., Mustill, A.~J., Bonsor, A., \& Wyatt, M.~C.\ 2013a, MNRAS, 431, 1686 

\bibitem[Veras et al.(2013b)]{veretal2013b} Veras, D., Hadjidemetriou, J.~D., \& Tout, C.~A.\ 2013b, MNRAS, 435, 2416 

\bibitem[Veras et al.(2014a)]{veretal2014a} Veras, D., Jacobson, S.~A., G\"{a}nsicke, B.~T.\ 2014a, MNRAS, 445, 2794 

\bibitem[Veras et al.(2014b)]{veretal2014b} Veras, D., Shannon, A., G\"{a}nsicke, B.~T.\ 2014b, MNRAS, 445, 4175

\bibitem[Veras et al.(2014c)]{veretal2014c} Veras, D., Evans, N.~W., Wyatt, M.~C., \& Tout, C.~A.\ 2014c, MNRAS, 437, 1127 

\bibitem[Veras et al.(2014d)]{veretal2014d} Veras, D., Leinhardt, Z.~M., Bonsor, A., G\"{a}nsicke, B.~T.\ 2014d, MNRAS, 445, 2244 

\bibitem[Veras et al.(2015a)]{veretal2015a} Veras, D., Eggl, S., G\"{a}nsicke, B.~T.\ 2015a, MNRAS, 451, 2814 

\bibitem[Veras et al.(2015b)]{veretal2015b} Veras, D., Eggl, S., G\"{a}nsicke, B.~T.\ 2015b, MNRAS, 452, 1945 

\bibitem[Veras \& G\"{a}nsicke(2015)]{vergae2015} Veras, D., G\"{a}nsicke, B.~T.\ 2015, MNRAS, 447, 1049 

\bibitem[Veras(2016)]{veras2016} Veras, D., 2016, Royal Society Open Science, 3:150571.

\bibitem[Veras et al.(2016)]{veretal2016} Veras, D., Marsh, T.~R., G\"{a}nsicke, B.~T.\ 2016, Submitted to MNRAS

\bibitem[Villaver et al.(2014)]{viletal2014} Villaver, E., Livio, M., Mustill, A.~J., \& Siess, L.\ 2014, ApJ, 794, 3 

\bibitem[Voyatzis et al.(2013)]{voyetal2013} Voyatzis, G., Hadjidemetriou, J.~D., Veras, D., \& Varvoglis, H.\ 2013, MNRAS, 430, 3383

\bibitem[Walsh et al.(2011)]{waletal2011} Walsh, K.~J., Morbidelli, A., Raymond, S.~N., O'Brien, D.~P., \& Mandell, A.~M.\ 2011, Nature, 475, 206 

\bibitem[Winn \& Fabrycky(2015)]{winfab2015} Winn, J.~N., \& Fabrycky, D.~C.\ 2015, ARA\&A, 53, 409 

\bibitem[Wyatt et al.(2014)]{wyaetal2014} Wyatt, M.~C., Farihi, J., Pringle, J.~E., \& Bonsor, A.\ 2014, MNRAS, 439, 3371 

\bibitem[Xu \& Jura(2012)]{xujur2012} Xu, S., \& Jura, M.\ 2012, ApJ, 745, 88 

\bibitem[Xu et al.(2016)]{xuetal2016} Xu, S., Jura, M., Dufour, P., \& Zuckerman, B.\ 2016, In Press ApJL, arXiv:1511.05973 

\bibitem[Zakamska \& Tremaine(2004)]{zaktre2004} Zakamska, N.~L., \& Tremaine, S.\ 2004, AJ, 128, 869 

\bibitem[Zuckerman et al.(2003)]{zucetal2003} Zuckerman, B., Koester, D., Reid, I.~N., H\"{u}nsch, M.\ 2003, ApJ, 596, 477 

\bibitem[Zuckerman et al.(2010)]{zucetal2010} Zuckerman, B., Melis, C., Klein, B., Koester, D., \& Jura, M.\ 2010, ApJ, 722, 725 

\end{thebibliography}
\end{document}